\documentclass[extra, referee]{gji}

\usepackage{timet}
\usepackage{url} 
\usepackage{lineno}
\usepackage{soul}
\usepackage{nccmath}
\usepackage{enumitem}
\usepackage{ifthen}
\usepackage{color}
\usepackage{xcolor}
\usepackage{graphicx}
\usepackage{amsmath}
\usepackage{amssymb}
\usepackage{algorithm}
\usepackage{algorithmic}
\usepackage{setspace}
\usepackage{multirow}
\usepackage[T1]{fontenc}
\usepackage{makecell}

\DeclareMathOperator*{\argmax}{argmax}
\graphicspath{{./Fig/}}
\title[Bayesian Geophysical Inversion using SMVI]{Bayesian Geophysical Inversion using Stein Mixture Variational Inference for Multimodal or Complex Solutions}
\author[Zhao and Curtis, 2026]
{Xuebin Zhao$^1$ and Andrew Curtis$^{1}$\\
	$^1$ School of GeoSciences, University of Edinburgh, Edinburgh, United Kingdom \\
	E-mail: \textit{xuebin.zhao@ed.ac.uk, andrew.curtis@ed.ac.uk}
}

\begin{document}
\label{firstpage}


\begin{titlepage}
	\begin{center}
		\vspace*{1cm} 
		
		\huge
		\textbf{Bayesian Geophysical Inversion using Stein Mixture Variational Inference for Multimodal or Complex Solutions}
		
		\vspace{1.0cm}
		\LARGE
		Xuebin Zhao$^{1}$ and Andrew Curtis$^{1}$\\
		
		\vspace{0.5cm}
		\Large
		School of GeoSciences, University of Edinburgh, United Kingdom \\
		
		\vspace{0.4cm}
		\Large
		E-mail: \textit{xuebin.zhao@ed.ac.uk, andrew.curtis@ed.ac.uk}
		
		\vfill
		\vfill
	\end{center}
\end{titlepage}

\newpage

\begin{summary}
Bayesian geophysical inversion provides a probabilistic framework to estimate subsurface properties, in which uncertainties are represented by a so-called posterior probability density function (pdf). Estimating this pdf is computationally challenging when it is high dimensional or complex, particularly if it may be multimodal. We introduce Stein mixture variational inference (SMVI) for efficient Bayesian geophysical inversion with multimodal or complex posterior solutions. The method combines parametric variational inference with Stein variational gradient descent (SVGD). In SMVI, the posterior pdf is approximated by a mixture (sum) of distributions, in which component distributions are optimised jointly using the SVGD update rule. This construction combines computational efficiency with the flexibility to represent a range of plausible Earth models and allows components to be updated in parallel. Here, SMVI is first tested on two synthetic seismic inverse problems, demonstrating that SMVI can recover non-Gaussian, complex and multimodal posterior solutions accurately and efficiently. Then, we apply SMVI to real seismic data by performing Love wave tomography of the British Isles, and obtain inversion results that are consistent with previous Monte Carlo solutions. Around the East Irish Sea sedimentary basins, SMVI captures bimodal uncertainty associated with ambiguity in basin boundary locations, revealing alternative subsurface structures compatible with the observations. These results indicate that SMVI generalises various previous variational methods, and provides a practical and effective framework for quantifying complex uncertainties and supporting geological interpretation of seismic images.
\end{summary}


\newpage

\section{Introduction}
Geophysical investigations to estimate subsurface properties using indirect observations usually require an inverse problem to be solved. A so-called forward function is first defined, which predicts data that would be observed given a set of model parameters that represent the Earth's properties. The aim of inversion is then to estimate or characterise the family of parameter values that are consistent with a specific set of real-world observations or measurements, and with any other information or constraints that are known \textit{a priori} \citep{tarantola2005inverse}.

In practice, solutions to geophysical inverse problems are non-unique because observed data are finite and noisy, subsurface illumination is incomplete, acquisition geometries are limited as observations are normally available only near the surface, and the forward function is often nonlinear and is only an approximation to the true physics \citep{mosegaard2002monte, sambridge2006trans, valentine2023emerging, fichtner2025high}. Trade-offs among model parameters further complicate the interpretation of the inversion results so that only certain questions about the subsurface can be answered with confidence \citep{arnold2018interrogation}. It is therefore important to estimate uncertainties in inverse solutions accurately, both to describe the range of plausible Earth models, and to support subsequent subsurface interpretation and risk-based decision-making \citep{poliannikov2016effect, arnold2018interrogation, ely2018assessing, siahkoohi2022deep, zhao2026linearised}.

Bayesian inference solves inverse problems within a probabilistic framework by using Bayes rule to update prior information about model parameters with new information from observed data, to calculate a so-called \textit{posterior} probability distribution function (pdf). This describes the family of Earth models consistent with both the data and the prior information, and assigns relative probabilities to different model parameter values. Bayesian inference therefore provides a natural means of estimating uncertainty, parameter trade-offs, and the range of admissible subsurface structures.

For strongly nonlinear inverse problems, posterior pdf's can be highly non-Gaussian, exhibit strong and nonlinear correlations, and in some cases are multimodal, meaning that multiple distinct regions of parameter space provide similarly acceptable fits to the data and prior information \citep{tarantola2005inverse, sambridge2014parallel, sambridge2025trans}. In geophysical applications, such behaviour can arise from cycle skipping in full waveform inversion \citep{hoffmann2024local}, multipathing in wave propagation \citep{zhang2023surface}, trade-offs between subsurface structure and physical properties \citep{gebraad2020bayesian}, and sparse or band-limited data coverage \citep{galetti2015uncertainty}. Methods used to solve such inverse problems should therefore be able to represent complex posterior pdf's when the need arises.

Bayesian inversion is often solved using Monte Carlo random sampling methods. A typical group of such methods is Markov chain Monte Carlo (McMC), in which an ensemble of model realisations is drawn from the posterior distribution \citep{metropolis1953equation, hastings1970monte, mosegaard1995monte, sambridge2002monte, bodin2009seismic, fichtner2019hamiltonian, khoshkholgh2021informed, zhang2023surface, magrini2025bayesbay}. These samples can then be used to approximate statistical properties of the posterior distribution. However, such methods are often computationally expensive for large, high dimensional inverse problems, because the number of samples required to characterise the posterior pdf increases exponentially with the model space (dimensionality the number of unknown parameters) owing to the curse of dimensionality \citep{curtis2001prior}. In addition, convergence can be slow when the posterior distribution is strongly correlated or multimodal \citep{gelman1992inference}.

Variational inference provides an efficient alternative to Monte Carlo sampling for Bayesian inversion. In variational methods, a family of known and computationally tractable probability distributions, referred to as the variational family, is defined in advance. An optimal member of this family, referred to as the variational distribution, is then chosen to approximate the true posterior pdf, by minimising the difference between the variational and posterior distributions \citep{blei2017variational, zhang2018advances}. In this way, variational methods replace repeated stochastic sampling with numerical optimisation while still providing a probabilistic characterisation of uncertainty. 

This optimisation-based viewpoint is particularly appealing for geophysical applications in which forward modelling is expensive and scalability is a primary concern \citep{zhang20233, lomas20233d, chen2025variational, siahkoohi2026dual}. As a result, variational methods have been developed and applied to a range of geophysical inverse problems, including petrophysical and lithological inversion \citep{nawaz2018variational, nawaz2019rapid}, travel time tomography \citep{zhang2019seismic, zhao2021bayesian, levy2022variational}, seismic migration \citep{siahkoohi2022deep}, surface wave inversion \citep{zhao2025bayesian, yang20253dswi}, full waveform inversion \citep{zhang2021bayesianfwi, bates2022probabilistic, wang2023re, zhao2024physically, yin2024wise, berti2025bayesian, siahkoohi2026dual}, time-lapse imaging \citep{zhang2024bayesian, sun2024enabling, zhao2025uncertainty, li2025uncertainty}, and experimental design \citep{strutz2023variational}.

Variational methods can broadly be divided into two main types: parametric variational inference methods and sample-based nonparametric methods. Parametric variational methods approximate the shape of the posterior pdf using an explicit mathematical expression parametrised by a set of optimisable control parameters \citep{rezende2015variational, guo2016boosting, kucukelbir2017automatic, sjolund2023tutorial, zhao2024physically}, and many such methods are developed for geophysical inverse problems within the \textit{GeoPVI} code package \citep{zhao2026geopvi}. Among these, automatic differentiation variational inference \cite[ADVI --][]{kucukelbir2017automatic} and physically structured variational inference \cite[PSVI --][]{zhao2024physically}, both based on transformed and optimised Gaussian variational distributions, have proved especially practical and computationally efficient for high dimensional, large scale inverse problems such as full waveform inversion.

The second class is sample- or particle-based variational methods, exemplified by Stein variational gradient descent \cite[SVGD --][]{liu2016stein}. Rather than optimising a set of hyper-parameters that defines a parametric variational distribution, SVGD iteratively updates an ensemble of samples, in this context called particles because they move, according to a deterministic transport rule derived from Stein's method and kernel interactions \citep{liu2016kernelized}. The final particle ensemble is then used to approximate the posterior distribution. This provides a flexible, sample-based, nonparametric approximation and can represent posterior pdf's that differ substantially from unimodal, Gaussian-like distributions.

Parametric and particle-based variational methods offer complementary advantages, but each also has important limitations. Parametric methods are efficient, and are easier to scale to large scale geophysical inverse problems, but their performance depends on the expressiveness of the chosen variational family \citep{zhao2026geopvi}. In particular, a single transformed Gaussian approximation, as used in ADVI and PSVI, can not represent strongly non-Gaussian or multimodal posterior distributions accurately \citep{ronning2025elboing, yang20253dswi}. Particle-based methods such as SVGD and its variants \cite[e.g.,][]{gallego2018stochastic, wang2019nonlinear, d2021annealed} are more flexible and in principle can adapt more naturally to complicated posterior distributions, but this flexibility commonly comes with a significantly increased computational cost \cite[see comparison in][]{zhao2024bayesian}. In particular, the same number of particles or samples required at the end of the inversion must be carried through at least part of the optimisation, and this number is typically large if we are to characterise high-dimensional pdf's accurately. In practise, particle ensembles may fail to populate all relevant posterior modes because of the limited number of particles that can be used in high dimensional problems, leading to a problem commonly referred to as \textit{particle degeneracy} or \textit{variance collapse} \citep{d2021annealed, ba2021understanding, zhang20233}.

In this paper, we introduce \textit{Stein mixture variational inference} (SMVI) for Bayesian inverse problems that may have multimodal or complex posterior distributions. The method uses parametric variational inference in a manner that mirrors the derivation of SVGD. In SMVI, the posterior pdf is approximated using a mixture variational distribution, in which each component of the mixture is represented by a parametric variational family defined by a finite set of hyper-parameters. Each component is also denoted as a particle herein, and an example of a particle would be a Gaussian distribution. Different particles parametrise different components of posteriori distributions, and are optimised jointly using the SVGD update rule \citep{ronning2025elboing}. In this way, SMVI retains a tractable parametric description within each component, and uses a mixture of such components to represent an approximate posterior distribution that can be as complex as desired. 

In SVGD, each particle represents a single model realisation and therefore provides information only at a single point in parameter space. By contrast, in SMVI each particle defines a parametric component distribution and therefore carries information within at least a local neighbourhood of parameter space. This gives the mixture approximation greater flexibility and expressiveness than is possible using typical single parametric variational distributions, and retains more information than an SVGD particle ensemble with the same number of particles. As shown below, this property is useful to represent multimodal posterior distributions and can help to mitigate mode collapse in SVGD.

The SMVI framework is closely related to \textit{Boosting Variational Inference} (BVI), introduced to geophysics by \cite{zhao2024bayesian}, as both methods approximate the posterior pdf using a mixture variational distribution. The main distinction lies in how the mixture distribution is constructed and optimised. BVI adopts a sequential strategy, adding one component at a time and optimising each new component while keeping previously added components fixed. This strategy naturally discourages different components from converging to similar posterior modes.
By contrast, SMVI optimises all mixture components simultaneously. Instead of relying on sequential boosting to separate components, SMVI introduces entropy regularisation, implemented through the repulsive interaction inherited from SVGD, to encourage diversity among components during optimisation. As we show in a full waveform inversion example below, simultaneous optimisation is attractive for large scale geophysical inverse problems because all components can in principle be updated in parallel, allowing effective use of modern high performance computing (HPC) architectures and reducing overall wall-clock time when sufficient computational resources are available. 

The rest of this paper is organised as follows. We first review variational Bayesian inversion, ADVI, PSVI and SVGD. We then introduce the SMVI methodology and compare it with those former methods. In Section \ref{section:example}, the effectiveness of SMVI is demonstrated using three seismic inversion problems. A surface wave inversion example is used to assess the representation of multimodal posterior solutions. An acoustic full waveform inversion example is used to examine the computational efficiency and accuracy of the method. A Love wave tomography application beneath the British Isles then demonstrates the applicability of the method to real seismic data. Finally, we provide a brief discussion and conclude.

\section{Methodology}
\subsection{Variational Bayesian Inference}
In Bayesian inference, the solution to an inverse problem is represented by the so-called \textit{posterior} probability density function (pdf) $p(\mathbf{m}|\mathbf{d}_{obs})$ which describes more or less probable values of a model vector $\mathbf{m}$ given observed data $\mathbf{d}_{obs}$, and is calculated using Bayes' theorem
\begin{equation}
	p(\mathbf{m}|\mathbf{d}_{obs}) = \dfrac{p(\mathbf{d}_{obs}|\mathbf{m})p(\mathbf{m})}{p(\mathbf{d}_{obs})}
	\label{eq:bayes}
\end{equation}
where $p(\mathbf{m})$ is the prior pdf of $\mathbf{m}$, which describes the \textit{a priori} or prior knowledge about $\mathbf{m}$ known independently of $\mathbf{d}_{obs}$, and $p(\mathbf{d}_{obs}|\mathbf{m})$ is the likelihood function, which denotes the probability of observing $\mathbf{d}_{obs}$ given a particular Earth model $\mathbf{m}$ (i.e., under the assumption that $\mathbf{m}$ represents the true state of the Earth). The term $p(\mathbf{d}_{obs})$ is called the \textit{evidence} and acts as a normalisation constant; it is equal to the integral of the numerator over all of parameter space, which is normally computationally intractable in high-dimensional problems.

In variational inference, a family of known probability distributions is defined (denoted as $\mathcal{Q} = \{q(\mathbf{m})\}$), from which an optimal variational distribution $q^*(\mathbf{m})$ is selected to best approximate the unknown posterior distribution. This optimal member can be obtained by minimising the Kullback-Leibler divergence \cite[KL-divergence,][]{kullback1951information} between the variational and posterior distributions:
\begin{equation}
\text{KL}[q(\mathbf{m})||p(\mathbf{m}|\mathbf{d}_{obs})] = \mathbb{E}_{q(\mathbf{m})}[\log q(\mathbf{m}) - \log p(\mathbf{m}|\mathbf{d}_{obs})]
\label{eq:kl}
\end{equation}
The KL divergence is a measure of difference between two pdf's; it is non-negative and equals zero only when the two distributions are identical. Minimising the KL-divergence is computationally intractable, since it requires $p(\mathbf{m}|\mathbf{d}_{obs})$ (and thus the evidence term $p(\mathbf{d}_{obs})$) to be evaluated. By substituting equation \ref{eq:bayes} into equation \ref{eq:kl}, we find that minimising the KL-divergence is achieved by maximising an evidence lower bound (ELBO) of $\log p(\mathbf{d}_{obs})$ \citep{bishop2006pattern, blei2017variational, zhang2018advances}:
\begin{equation}
	\text{ELBO}[q(\mathbf{m})] = \mathbb{E}_{q(\mathbf{m})}[\log p(\mathbf{m}, \mathbf{d}_{obs}) - \log q(\mathbf{m})]
	\label{eq:elbo}
\end{equation}
using gradient-based optimisation algorithms. Note that by definition of $\mathcal{Q}$ the optimisation result is then a valid probability distribution rather than a single model solution, and is therefore implicitly normalised by the correct evidence.

Different variational algorithms make different choices for the variational family. Below, we first review two types of methods that have been tested in a wide range of geophysical problems. We then introduce a new algorithm that combines properties of these two types of methods.

\subsection{Automatic Differentiation Variational Inference (ADVI) and Physically Structured Variational Inference (PSVI)}
\label{section:psvi}
Automatic differentiation variational inference \cite[ADVI,][]{kucukelbir2017automatic} is an efficient variational method that uses a transformed Gaussian distribution to approximate the posterior pdf. A Gaussian distribution $\mathcal{N}(\boldsymbol\eta; \boldsymbol\mu, \boldsymbol\Sigma)$ for a random variable $\boldsymbol\eta$ is defined uniquely by a mean vector $\boldsymbol\mu$ and a covariance matrix $\boldsymbol\Sigma$. Here, the Gaussian distribution is defined for a random vector $\boldsymbol\eta$ in the space of real numbers (i.e., an unconstrained space), whereas geophysical model vector parameters often have hard physical bounds (e.g., seismic velocity values typically lie within a range of $(0, 10)$ km/s). A bounded transform is therefore applied to map $\boldsymbol\eta$ to a constrained random variable $\mathbf{m}$, ensuring that the transformed vector satisfies the required physical constraints:
\begin{equation}
	\begin{split}
		& m_i = T(\eta_i) = a_i + \dfrac{b_i - a_i}{1+\exp (-\eta_i)}\\
		& \eta_i = T^{-1}(m_i) = \log(m_i - a_i) - \log(b_i - m_i) 
	\end{split}
	\label{eq:log_transform}
\end{equation}
where $\eta_i$ is the $i$-th element of the Gaussian vector $\boldsymbol\eta$ defined in an unconstrained space, and $m_i$ is the transformed geophysical parameter bounded between lower and upper bounds $a_i$ and $b_i$, respectively. Note that $a_i$ and $b_i$ are pre-defined and fixed during variational inversion, and $q(\mathbf{m})$ can be calculated using the change of variable formula
\begin{equation}
\log q(\mathbf{m}) = \log \mathcal{N}(\boldsymbol{\eta}; \boldsymbol\mu, \boldsymbol\Sigma) - \log|\det (\partial_{\boldsymbol{\eta}} T(\boldsymbol{\eta}))|
\label{eq:change_variable}
\end{equation}
where $|\det(\partial_{\boldsymbol{\eta}} T(\boldsymbol{\eta}))|$ calculates the absolute value of the determinant of the Jacobian matrix $\partial_{\boldsymbol{\eta}} T(\boldsymbol{\eta})$ \citep{rezende2015variational}. In this way, the Gaussian distribution can be optimised in the unconstrained $\boldsymbol\eta$-space, while the corresponding variational distribution $q(\mathbf{m})$ is induced in the physical model space through the transform in equations \ref{eq:log_transform} and \ref{eq:change_variable}. ADVI therefore approximates the posterior pdf using a transformed Gaussian distribution in the space of parameter $\mathbf{m}$.

In ADVI, the Gaussian parameters $\boldsymbol\mu$ and $\boldsymbol\Sigma$ in the unconstrained $\boldsymbol\eta$-space are optimised by maximising ELBO$[q(\mathbf{m})]$ (equation \ref{eq:elbo}) using a gradient-based optimisation algorithm. Evaluation of ELBO$[q(\mathbf{m})]$ and its gradient involves calculating the determinant $|\boldsymbol\Sigma|$ which can be prohibitively expensive for high dimensional inverse problems \citep{kucukelbir2017automatic, zhao2024physically}. A Cholesky decomposition is often introduced to reparametrise $\boldsymbol\Sigma$ with $\boldsymbol\Sigma = \mathbf{L}\mathbf{L^T}$, where $\mathbf{L}$ is a lower triangular matrix. The determinants of $\mathbf{L}$ and hence $\boldsymbol\Sigma$ can then be calculated easily.

Different parametrisations of $\mathbf{L}$ result in different algorithms. \cite{kucukelbir2017automatic} introduced \textit{full rank ADVI}, in which a dense matrix $\mathbf{L}$, and thus $\boldsymbol\Sigma$, is constructed and optimised to capture correlation information between model parameters. However, for a $k$-dimensional inverse problem, $\mathbf{L}$ requires $k(k+1)/2$ elements to be optimised, which can be memory intensive and computationally intractable for large $k$. 

A mean field approximation (i.e., \textit{mean field ADVI}) is introduced to reduce memory requirements, in which both $\mathbf{L}$ and $\boldsymbol\Sigma$ are set to be diagonal matrices \citep{kucukelbir2017automatic}. While computationally efficient, the method ignores correlations among different pairs of model parameters and has been shown to provide underestimated and biased uncertainties in various geophysical problems \citep{zhang20233, zhao2024bayesian}.

\cite{zhao2024physically} proposed \textit{physically structured variational inference} (PSVI), which parametrises matrix $\mathbf{L}$ using a sparse, physics-guided structure. Instead of modelling all parameter correlations, PSVI models only correlations that are expected \textit{a priori} to be significant in $\mathbf{m}$, for example between pairs of parameters that are in spatial proximity to one another, while setting all other parameter correlations to zero. In this way, PSVI retains the ability to capture dominant posterior correlations compared with mean field ADVI, while significantly reducing computational and memory requirements compared with full rank ADVI. PSVI has been shown to provide reasonably accurate posterior statistics, especially in spatially 3D geophysical inverse problems \citep{zhao2025efficient, zhao2025bayesian, yang20253dswi}. \cite{zhao2024bayesian} extended ADVI methods by gradient boosting: sequentially adding additional Gaussian distributions to the posterior pdf approximation, so as to further reduce the ELBO. 

Regardless of different methods used to parametrise $\mathbf{L}$, the objective is to maximise ELBO$[q(\mathbf{m})]$ with respect to $\mathbf\Theta = [\boldsymbol\mu, \mathbf{L}]$ iteratively using
\begin{equation}
	\mathbf\Theta^{l+1} = \mathbf\Theta^{l} + \epsilon^l \nabla_{\mathbf\Theta} \text{ELBO}[q(\mathbf{m}; \mathbf\Theta)]
	\label{eq:advi_update_theta}
\end{equation}
where $\epsilon^l$ is the step size at the $l$-th iteration. Here, the explicit dependence of the variational distribution $q(\mathbf{m}; \mathbf\Theta)$ on $\mathbf\Theta$ is written to highlight that $q(\mathbf{m}; \mathbf\Theta)$ is parametrised by $\mathbf\Theta$. Calculation of $\nabla_{\mathbf\Theta} \text{ELBO}[q(\mathbf{m}; \mathbf\Theta)]$ is performed using a Monte Carlo approximation. Details can be found in \cite{zhao2024physically} and in equation \ref{eq:grad_elbo} in Section \ref{section:smvi}.

\subsection{Stein Variational Gradient Descent (SVGD)}
\label{section:svgd}
In Stein variational gradient descent \cite[SVGD,][]{liu2016stein} the variational distribution $q(\mathbf{m})$ is represented implicitly by a set of $n$ model samples (particles) $\{\mathbf{m}_i\}_{i=1}^n$ (here subscript $i$ denotes the particle index). These particles are generated from an initial distribution (e.g., the Bayesian prior distribution), and are then optimised such that their empirical distribution approximates the posterior distribution. This is achieved by minimising the KL-divergence between the posterior and variational pdf's iteratively within a reproducing kernel Hilbert space \citep{liu2016kernelized}. Within each iteration, this is accomplished by perturbing each particle along a direction $\mathbf\Phi(\mathbf{m})$:
\begin{equation}
	\mathbf{m}_i^{l+1} = \mathbf{m}_i^{l} + \epsilon^l \mathbf\Phi^*_l(\mathbf{m}_i^l)
	\label{eq:svgd_update}
\end{equation}
The optimal perturbation $\mathbf\Phi^*(\mathbf{m})$ which defines the fastest direction that minimises the KL-divergence is given by \citep{liu2016stein}
\begin{equation}
	\mathbf\Phi^*_l(\mathbf{m}) = \frac{1}{n}\sum_{j=1}^{n}\left[k(\mathbf{m}^l_j, \mathbf{m}) \nabla_{\mathbf{m}_j^l}\log p(\mathbf{m}_j^l|\mathbf{d}_{obs}) + \nabla_{\mathbf{m}_j^l} k(\mathbf{m}^l_j, \mathbf{m}) \right]
	\label{eq:svgd_gradient}
\end{equation}
where $n$ is the number of particles and $l$ is the iteration index. Term $\nabla_{\mathbf{m}}\log p(\mathbf{m}|\mathbf{d}_{obs})$ is the gradient of the logarithmic posterior probability density $\log p(\mathbf{m}|\mathbf{d}_{obs})$ with respect to $\mathbf{m}$. In geophysical problems, this term can be calculated using the Jacobian information of a forward function \citep{zhang2019seismic}. The term $k(\mathbf{m}', \mathbf{m})$ is called a kernel function, and in this work we employ the commonly used radial basis function \cite[RBF,][]{liu2016stein}
\begin{equation}
	k(\mathbf{m}', \mathbf{m}) = \exp \left[-\frac{1}{h}||\mathbf{m}' - \mathbf{m}||^2\right]
	\label{eq:rbf_kernel}
\end{equation}
where $h$ is a factor that scales the intensity of interaction between different particles based on the distance between two particles $||\mathbf{m}' - \mathbf{m}||^2$. Following previous studies \citep{liu2016stein, zhang2019seismic}, we define $h=\tilde{d}/\sqrt{2\log n}$, where $\tilde{d}$ is the median of the pairwise distances between all pairs of particles. Term $\nabla_{\mathbf{m}'} k(\mathbf{m}', \mathbf{m})$ in equation \ref{eq:svgd_gradient} calculates the gradient of the kernel function with respect to a model sample $\mathbf{m}'$, and acts as a \textit{repulsive force} that prevents all model particles from converging to similar locations in parameter space (i.e., collapsing towards a single mode). In contrast, the first term in equation \ref{eq:svgd_gradient} consists of kernel-weighted gradients that move the particles towards high posterior probability regions. The combination of both terms ensures that particles converge to positions in which sample density approximates the posterior pdf -- given a sufficiently large number of particles, and sufficiently small step size in equation \ref{eq:svgd_update} \citep{liu2016stein}. 

For notational simplicity, equations \ref{eq:svgd_update}-\ref{eq:rbf_kernel} are written in terms of model vector $\mathbf{m}$. In geophysical applications, however, model parameters commonly have hard physical bounds, so SVGD is actually implemented in the unconstrained transformed space defined by equation \ref{eq:log_transform}. That is, particles are first mapped from the bounded model parameter space into the corresponding unconstrained space, and the SVGD update is performed in this space. After optimisation, the particles are transformed back to the original bounded physical space to approximate the posterior distribution. In this way, the optimisation remains consistent with the SVGD formulation while ensuring that all model parameters satisfy their physical constraints. This transformed space implementation has been used in previous geophysical applications of SVGD to ensure that bounded model parameters remain physically admissible during optimisation \citep{zhang2019seismic, zhang20233}.

\subsection{Stein Mixture Variational Inference (SMVI)}
\label{section:smvi}
While previous studies have shown that PSVI is an efficient variational algorithm \cite[e.g.,][]{zhao2025efficient, yang20253dswi}, the method assumes a transformed Gaussian posterior distribution and therefore can not represent strongly multimodal posterior solutions accurately. By contrast, SVGD is a particle-based method that provides a flexible means of approximating multimodal posterior pdf's, but the algorithm is computationally expensive \citep{zhang20233, zhao2024bayesian} and can yield biased results because of the finite number of particles -- a problem referred to as \textit{variance collapse} in SVGD \citep{ba2021understanding}. In addition, standard SVGD yields an ensemble of particles rather than an explicit parametric density from which new posterior samples can be drawn directly. To obtain additional samples, one generally needs either to rerun the inversion, or to introduce a separate density estimation step \cite[e.g.,][]{dromigny2026approximation}, both of which can be expensive. 

We now introduce a new variational algorithm -- Stein mixture variational inference (SMVI) -- which combines a parametric variational family such as ADVI or PSVI, with a learning rule similar to SVGD \citep{ronning2025elboing}. In SMVI, we build a mixture variational distribution composed of $n$ component probability densities with equal weight coefficients
\begin{equation}
	q(\mathbf{m};\mathbf\Psi) = \dfrac{1}{n} \sum_{i=1}^{n} g_i(\mathbf{m}; \mathbf\Theta_i), \quad \text{with} \quad \mathbf\Psi = \{\mathbf\Theta_i\}_{i=1}^n
	\label{eq:mixture}
\end{equation}
where each $g_i(\mathbf{m}; \mathbf\Theta_i)$ represents a component distribution. Equal weights are assigned to each component for simplicity of optimisation (see below). Here, we use a parametric component distribution, in which an explicit formula defines its form and a control parameter $\mathbf\Theta_i$ parametrises it uniquely. For example, a Gaussian component distribution $g(\mathbf{m}; \mathbf\Theta) = \mathcal{N}(\mathbf{m}; \boldsymbol{\mu}, \mathbf\Sigma)$ is parametric and can be defined uniquely by its control parameter $\mathbf\Theta = [\boldsymbol\mu, \mathbf\Sigma]$. More examples of parametric component distributions can be found in \cite{zhao2026geopvi}. The mixture distribution $q(\mathbf{m};\mathbf\Psi)$ in equation \ref{eq:mixture} is thus also parametric and is defined by a parameter set $\mathbf\Psi = \{\mathbf\Theta_i\}_{i=1}^n$, consisting of $n$ parameters $\mathbf\Theta_i$, each parametrising a component distribution $g_i(\mathbf{m}; \mathbf\Theta_i)$. Later in this paper, each $\mathbf\Theta_i$ is also referred to as a particle because of the way it is optimised. Therefore in SMVI, each particle $\mathbf\Theta_i$ does not represent a model sample in parameter space as in SVGD, but a complete, valid probability distribution, and the particle ensemble $\mathbf\Psi = \{\mathbf\Theta_i\}_{i=1}^n$ represents a collection of component distributions rather than a collection of pointwise model realisations.

The mixture distribution is then used to approximate the unknown posterior pdf by maximising the following ELBO$[q(\mathbf{m}); \mathbf\Psi]$ with respect to $\mathbf\Psi = \{\mathbf\Theta_i\}_{i=1}^n$
\begin{equation}
	\begin{split}
		\text{ELBO}[q(\mathbf{m}); \mathbf\Psi] & = \mathbb{E}_{q(\mathbf{m}; \mathbf\Psi)}[\log p(\mathbf{m}, \mathbf{d}_{obs}) - \log q(\mathbf{m}; \mathbf\Psi)] \\
		& = \dfrac{1}{n} \sum_{i=1}^{n} \mathbb{E}_{g_i(\mathbf{m}; \mathbf\Theta_i)}[\log p(\mathbf{m}, \mathbf{d}_{obs}) - \log q(\mathbf{m}; \mathbf\Psi)]
	\end{split}
	\label{eq:elbo_smvi}
\end{equation}
where the second row is obtained by substituting equation \ref{eq:mixture} into the first row. In this work, ADVI is used to define the component distributions $g_i(\mathbf{m}; \mathbf\Theta_i)$ for SMVI, and in principle the mixture distribution in equation \ref{eq:mixture} can approximate any continuous probability distribution, including the posterior pdf, to any desired level of accuracy \citep{bishop1994mixture, meier2007fully, earp2019probabilistic}.

Although it may appear straightforward to use a gradient-based optimisation algorithm to solve this problem, the non-convex nature of the objective function in equation \ref{eq:elbo_smvi} means that direct optimisation with a local search method carries the risk of converging to a degenerate solution, in which multiple particles $\mathbf\Theta_i$, and thus the corresponding component distributions $g_i(\mathbf{m}; \mathbf\Theta_i)$, converge to similar values. In such cases, the mixture variational distribution can not represent multimodal posterior solutions accurately.

This difficulty can be addressed within a variational optimisation framework generalised from the underlying principle of SVGD \citep{wang2019nonlinear}. Consider the following objective function to be maximised 
\begin{equation}
	F(\mathbf\Psi) = F(\{\mathbf{\Theta}_i\}_{i=1}^n) = \frac{1}{n}\sum_{i=1}^{n} t(\mathbf{\Theta}_i)
	\label{eq:F}
\end{equation}
for some function $t(\cdot)$. Note that the second row in equation \ref{eq:elbo_smvi} is a special case of equation \ref{eq:F}. Generally, $F(\{\mathbf{\Theta}_i\}_{i=1}^n)$ depends on $\{\mathbf{\Theta}_i\}_{i=1}^n$ through an empirical measure
\begin{equation}
	\rho = \rho(\mathbf{\Theta}) = \dfrac{1}{n} \sum_{i=1}^{n} \delta(\mathbf{\Theta} - \mathbf{\Theta}_i)
	\label{eq:rho}
\end{equation}
where $\delta(\mathbf{x})$ denotes the Dirac Delta function (unit impulse function), whose value is zero everywhere except at $\mathbf{x} = \mathbf{0}$, where it is infinite. Given this definition, $F(\{\mathbf{\Theta}_i\}_{i=1}^n)$ can be regarded as a \textit{functional} of $\rho(\mathbf{\Theta})$, which we denote by $F[\rho]$, i.e., $F(\mathbf\Psi) = F(\{\mathbf{\Theta}_i\}_{i=1}^n) = F[\rho]$. Here, $\rho(\mathbf{\Theta})$ should be interpreted as a continuous distribution over component parameters that is composed of a series of $n$ delta functions, and can be represented numerically by a finite particle set $\{\mathbf{\Theta}_i\}_{i=1}^n$.

The maximisation of $F[\rho]$ with respect to $\rho$ is non-convex and tends to converge towards a similarly degenerate solution as discussed above. This can be avoided by solving the following entropy regularised optimisation problem \citep{wang2019nonlinear}
\begin{equation}
\rho^* = \argmax_{\rho} \left( F[\rho] + \alpha\mathbb{H}[\rho]\right)
\label{eq:F_regularised}
\end{equation}
where $\mathbb{H}[\rho] = -\int \rho(\mathbf{\Theta})\log \rho(\mathbf{\Theta}) d\mathbf{\Theta}$ or $ -\sum_{i=1}^n \rho(\mathbf{\Theta}_i) \log \rho(\mathbf{\Theta}_i)$, depending on whether $\rho$ is continuous or discretised, measures the entropy -- i.e., the uncertainty represented by $\rho$. This entropy term regularises the continuous parameter distribution $\rho(\mathbf{\Theta})$, while the particles $\{\mathbf{\Theta}_i\}_{i=1}^n$ provide a discrete approximation to it. In this way, the regularised optimisation is formulated at the level of a continuous distribution, even though it is implemented using a finite set of particles. The regularised formulation in equation \ref{eq:F_regularised} therefore encourages $\rho$ to be more diffuse, thereby preventing the particles $\{\mathbf{\Theta}_i\}_{i=1}^n$ from converging to identical values. Parameter $\alpha$ is a regularisation factor that controls the weight of the entropy term.

\cite{wang2019nonlinear} extended the key idea from SVGD, and derived an efficient particle-based method to solve the optimisation problem in equation \ref{eq:F_regularised}. In particular, a set of $n$ particles $\{\mathbf{\Theta}_i\}_{i=1}^n$ that defines $\rho$ is initialised and then perturbed iteratively to maximise equation \ref{eq:F_regularised}, following a functional gradient direction constrained within a reproducing kernel Hilbert space (RKHS), similarly to equation \ref{eq:svgd_update}
\begin{equation}
\mathbf{\Theta}_i^{l+1} = \mathbf{\Theta}_i^{l} + \epsilon^l \mathbf\Phi^*_l(\mathbf{\Theta}_i^l)
\label{eq:smvi_update}
\end{equation}
where $l$ denotes the iteration index, and $\epsilon^l$ is a small step size. The optimal $\mathbf\Phi^*(\mathbf{\Theta})$ that defines the fastest direction to maximise equation \ref{eq:F_regularised} can be derived within the RKHS
\begin{equation}
	\mathbf\Phi^*_l(\mathbf{\Theta}) = \sum_{j=1}^{n}\left[k(\mathbf{\Theta}^l_j, \mathbf{\Theta}) \nabla_{\mathbf{\Theta}_j^l} F(\mathbf\Psi) + \frac{\alpha}{n} \nabla_{\mathbf{\Theta}_j^l} k(\mathbf{\Theta}^l_j, \mathbf{\Theta}) \right]
	\label{eq:nsvgd_gradient}
\end{equation}
This equation takes a similar form to that in equation \ref{eq:svgd_gradient} used in SVGD: term $\nabla_{\mathbf{\Theta}} F(\mathbf\Psi)$ calculates the gradient of $F(\mathbf\Psi)$ (equation \ref{eq:F}, without the entropy regularisation term) with respect to particle $\mathbf{\Theta}$; term $k(\mathbf{\Theta}', \mathbf{\Theta})$ is a kernel function following the definition in equation \ref{eq:rbf_kernel}; and term $\nabla_{\mathbf{\Theta}'} k(\mathbf{\Theta}', \mathbf{\Theta})$ represents the corresponding repulsive force that prevents all particles $\{\mathbf{\Theta}_i\}_{i=1}^n$ from converging to similar values. Therefore, the first term in equation \ref{eq:nsvgd_gradient} drives $\{\mathbf{\Theta}_i\}_{i=1}^n$ so as to increase $F(\mathbf\Psi)$ using the gradient information $\nabla_{\mathbf{\Theta}} F(\mathbf\Psi)$, while the second term is used to maximise the entropy $\mathbb{H}[\rho]$, thereby promoting diversity among $\{\mathbf{\Theta}_i\}_{i=1}^n$. The value of $\alpha$ controls the magnitude of the repulsive force. The output of the above algorithm is a finite set of particles that approximately maximises the regularised objective function in equation \ref{eq:F_regularised}, subject to the finite particle representation \citep{wang2019nonlinear}.

Note that the algorithm described in equations \ref{eq:smvi_update} and \ref{eq:nsvgd_gradient} provides a general form for optimising mixture models. For example, SVGD described in Section \ref{section:svgd} can be viewed as a special case, in which $\alpha=1$, particles $\{\mathbf{\Theta}_i\}_{i=1}^n$ represent model samples $\{\mathbf{m}_i\}_{i=1}^n$, and $F(\mathbf\Psi) = \frac{1}{n}\sum_{i=1}^{n} \log p(\mathbf{m}_i|\mathbf{d}_{obs})$. 

For SMVI, the ELBO$[q(\mathbf{m}; \mathbf\Psi)]$ defined in equation \ref{eq:elbo_smvi} can be viewed as another special case of $F(\mathbf\Psi)$. Therefore, the above algorithm can be used to optimise $\mathbf\Psi^* = \{\mathbf{\Theta}_i^*\}_{i=1}^n$ that satisfies
\begin{equation}
\mathbf\Psi^* = \argmax_{\mathbf\Psi} \left( \text{ELBO}[q(\mathbf{m}; \mathbf\Psi)] + \alpha \mathbb{H}[\rho]\right)
\label{eq:regularised_elbo}
\end{equation}
However, as a variational algorithm, the primary target of SMVI is to maximise ELBO$[q(\mathbf{m}; \mathbf\Psi)]$. The regularised objective in equation \ref{eq:regularised_elbo} should therefore be interpreted as an entropy regularised surrogate that promotes diversity among component parameters while remaining connected to the standard ELBO. Nevertheless, it can be shown that for $\alpha = 1$, the regularised objective is itself a lower bound on the logarithmic evidence \citep{ronning2025elboing}. Under the conditions discussed therein, and in the infinite particle limit, optimisation of equation \ref{eq:regularised_elbo} maximises the standard ELBO formulation. Thus, in practice, SMVI maximises a regularised evidence lower bound whose optimum approaches that of ELBO$[q(\mathbf{m}; \mathbf\Psi)]$ as the number of particles increases.

The above analysis suggests the following algorithm to perform SMVI:
\begin{enumerate}
\item Initialise a set of particles $\mathbf\Psi^0 = \{\mathbf{\Theta}_i^0\}_{i=1}^n$ to parametrise an initial mixture variational distribution $q(\mathbf{m}; \mathbf\Psi^0)$.
\item At iteration $l$, update each $\mathbf{\Theta}_i$ using equation \ref{eq:smvi_update} where 
\begin{equation}
\mathbf\Phi^*_l(\mathbf{\Theta}) = \sum_{j=1}^{n}\left[k(\mathbf{\Theta}^l_j, \mathbf{\Theta}) \nabla_{\mathbf{\Theta}_j^l} \text{ELBO}[q(\mathbf{m}; \mathbf\Psi)] + \frac{1}{n} \nabla_{\mathbf{\Theta}_j^l} k(\mathbf{\Theta}^l_j, \mathbf{\Theta}) \right]
\label{eq:smvi_gradient}
\end{equation}
\item After convergence, use $\{\mathbf{\Theta}_i\}_{i=1}^n$ to retrieve the optimised variational distribution $q(\mathbf{m}; \mathbf\Psi)$.
\end{enumerate}

The only remaining issue is the calculation of derivatives $\nabla_{\mathbf{\Theta}_j} \text{ELBO}[q(\mathbf{m}; \mathbf\Psi)]$ with respect to particle $\mathbf{\Theta}_j$. \cite{ronning2025elboing} derived this term using a \textit{score function gradient estimator} for problems in which calculating the gradient $\nabla_\mathbf{m} \log p(\mathbf{m}, \mathbf{d}_{obs})$ is computationally expensive or numerically unstable, for example in reinforcement learning applications where the forward function is represented by neural networks \citep{mohamed2020monte}. In geophysical applications, physics-based gradients are available \cite[e.g., calculated using the adjoint-state method --][]{plessix2006review} and can be evaluated efficiently. In this work, we derive the following new gradient term for SMVI based on a \textit{reparametrisation trick} \citep{kingma2014auto}, which reduces the bias and variance of the gradient estimation and incorporates physics information explicitly:
\begin{equation}
	\begin{split}
	\nabla_{\mathbf{\Theta}_j} \text{ELBO}[q(\mathbf{m}; \mathbf\Psi)] & = \dfrac{1}{n} \sum_{i=1}^{n} \nabla_{\mathbf{\Theta}_j} \mathbb{E}_{g_i(\mathbf{m}; \mathbf\Theta_i)}[\log p(\mathbf{m}, \mathbf{d}_{obs}) - \log q(\mathbf{m}; \mathbf\Psi)] \\
	& = \dfrac{1}{n} \sum_{i=1}^{n} \mathbb{E}_{\mathcal{N}(\mathbf{0}, \mathbf{I})}[\nabla_{\mathbf{\Theta}_j} \log p(\mathbf{m}, \mathbf{d}_{obs}) - \nabla_{\mathbf{\Theta}_j} \log q(\mathbf{m}; \mathbf\Psi)] \\
	& \approx \dfrac{1}{nk} \sum_{i=1}^{n} \sum_{c=1}^{k} \left[\nabla_{\mathbf{\Theta}_j}\mathbf{m}_c^i \nabla_{\mathbf{m}_c^i}\log p(\mathbf{m}_c^i, \mathbf{d}_{obs}) - \nabla_{\mathbf{\Theta}_j} \log q(\mathbf{m}_c^i; \mathbf\Psi) \right]
	\end{split}
\label{eq:grad_elbo}
\end{equation}
The first line is obtained from equation \ref{eq:elbo_smvi}. In the second line, the expectation is taken with respect to a standard normal distribution $\mathcal{N}(\mathbf{0}, \mathbf{I})$ rather than a component distribution $g_i(\mathbf{m}; \mathbf\Theta_i)$ by reparametrisation \cite[derivation details can be found in Appendix A of][]{zhao2021bayesian}. In the third line, the expectation with respect to $\mathcal{N}(\mathbf{0}, \mathbf{I})$ is approximated by a Monte Carlo integration, using an average of the corresponding expression evaluated using $k$ samples $\{\mathbf{m}_c^i\}_{c=1}^{k}$ drawn from the $i$-th component distribution $g_i(\mathbf{m}; \mathbf\Theta_i)$ \citep{kucukelbir2017automatic}. To reduce the variance of the Monte Carlo integration, $\{\mathbf{m}_c^i\}_{c=1}^{k}$ are actually obtained by drawing $k$ samples $\{\boldsymbol{\xi}_c\}_{c=1}^{k}$ from $\mathcal{N}(\mathbf{0}, \mathbf{I})$ and then performing
\begin{equation}
	\begin{split}
	\mathbf{m}_c^i & = y(\boldsymbol{\xi}_c;\mathbf\Theta_i) \\
	& = T(\boldsymbol{\eta}_c^i) = T(\boldsymbol{\mu}_i + \mathbf{L}_i \boldsymbol{\xi}_c) 
	\end{split}
\label{eq:reparametrisation1}
\end{equation}
where $y(\cdot)$ is an invertible function that transforms a random sample $\boldsymbol{\xi}_c$ distributed according to $\mathcal{N}(\mathbf{0}, \mathbf{I})$ into sample $\mathbf{m}_c^i$ distributed according to $g_i(\mathbf{m}; \mathbf\Theta_i)$. For example, in this work since each component distribution is defined as an ADVI-type distribution, particle $\mathbf\Theta_i = [\boldsymbol{\mu}_i, \mathbf{L}_i]$ defines a Gaussian distribution, and thus gives the second line in equation \ref{eq:reparametrisation1}, where $T(\cdot)$ is defined in equation \ref{eq:log_transform} (i.e., transform a random variable from an unconstrained space into a constrained space). 


\subsection{Comparison of Different Methods}
\label{section:comparison}
In this section, we compare ADVI (including mean field ADVI, PSVI and full rank ADVI), SVGD, BVI and SMVI. SMVI can be viewed as a unifying variational framework: it connects parametric variational inference and particle-based Stein methods, while providing an attractive alternative to using boosting as a method to add complexity to solutions. In particular, ADVI-type methods correspond to the single component case of SMVI, in which the posterior approximation is described by one transformed Gaussian component. By introducing multiple component distributions, SMVI extends this representation and can therefore approximate multimodal posterior distributions more flexibly and accurately than a single ADVI or PSVI component.

SVGD can be interpreted as a limiting case of SMVI in which each component distribution takes the form of a Dirac Delta function with zero-volume support, i.e., $g_i(\mathbf{m}) = \delta(\mathbf{m} - \mathbf{m}_i)$, parametrised by a model particle $\mathbf{m}_i$. In this case, the variational approximation is represented by a set of Delta function spikes in parameter space
\begin{equation}
	q(\mathbf{m}) = \dfrac{1}{n}\sum_{i=1}^{n} \delta(\mathbf{m} - \mathbf{m}_i)
	\label{eq:smvi_svgd}
\end{equation}
and is parametrised by the particle locations. In contrast, in SMVI each particle parametrises a component probability distribution with non-zero support. Each particle therefore carries information about (at least) a local neighbourhood of continuous parameter space. This gives the approximation a smoother structure than a standard SVGD particle ensemble and helps to explain why SMVI can mitigate the variance collapse behaviour observed in finite particle SVGD solutions \citep{ba2021understanding, d2021annealed}. As shown in the toy example below and in Section \ref{section:example}, this property is useful for representing multimodal posterior distributions efficiently.

In both SMVI and BVI, a mixture variational distribution is optimised. SMVI combines ADVI and SVGD to construct a mixture distribution, whereas BVI combines ADVI and gradient boosting. In BVI, each component distribution is optimised and added to the mixture distribution sequentially, previously optimised components remaining fixed \citep{guo2016boosting, zhao2024bayesian}. In principle, any new BVI component is introduced to be different from, and to improve the accuracy of, the existing variational distribution. This provides a natural way to avoid multiple components converging towards similar solutions.

Similarly a sequential, boosting version of SMVI may also be constructed. Given an existing mixture of $n$ components, the parameters of those components could be fixed while an additional component is introduced and updated using the SMVI update rule in equation \ref{eq:smvi_gradient}. The update direction for the new component would be evaluated using the complete mixture distribution, invoking the usual Stein repulsive interaction with the previously obtained components. Although this restricted update would not generally be the optimal direction that maximally improves the variational objective (or minimises the KL-divergence) for \textit{all} components, it may still minimise the KL-divergence iteratively and yield an accurate posterior approximation after repeated component additions. In contrast to conventional BVI, such a formulation would retain the fixed, equal component weights used in equation \ref{eq:mixture}, while replacing the direct stochastic-gradient optimisation of a newly added BVI component with the RKHS-based SMVI (and SVGD) update. This approach may provide an adaptive way to construct an SMVI mixture and reduce the number of components required in some problems; however, it would introduce a sequential dependence between components, making parallelisation across component additions impossible.

\begin{figure}
	\centering\includegraphics[width=\textwidth]{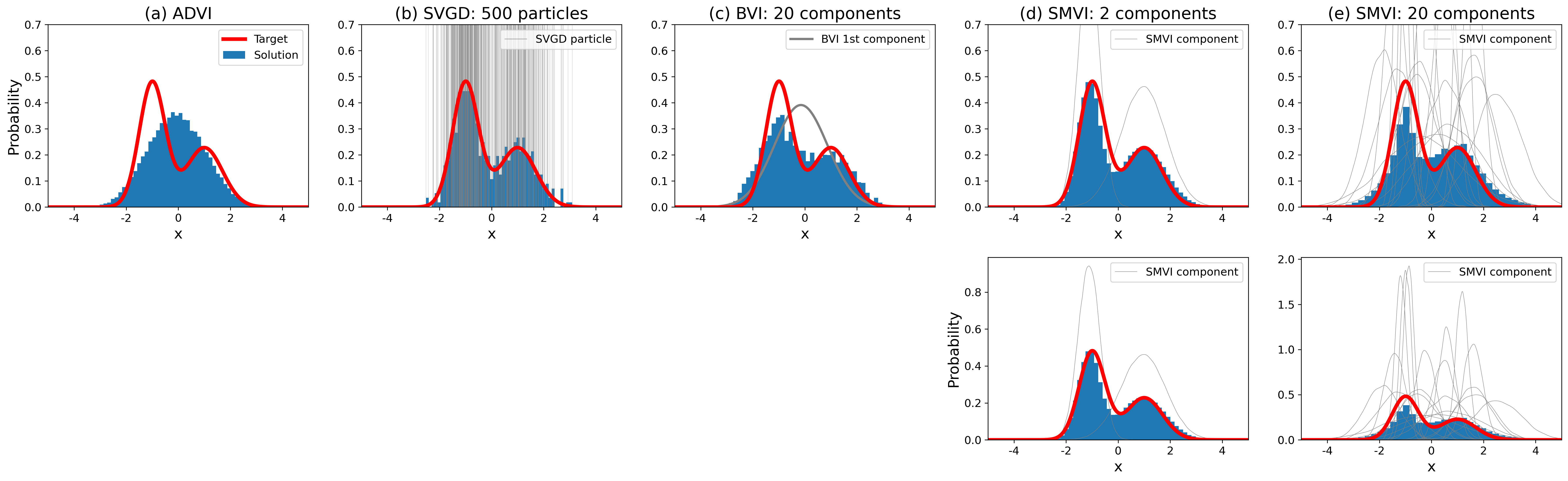}
	\caption{1D toy example. Panels (a), (b), (c), (d) and (e) show results obtained using ADVI, SVGD with 500 particles, BVI with 20 components, SMVI with 2 components, and SMVI with 20 components, respectively. In each panel, a red curve shows the target distribution, which is a mixture of two Gaussians, and a blue histogram represents the variational distribution obtained using the corresponding method. Grey vertical lines in (b) show the locations of the 500 particles. A grey curve in (c) shows the first component optimised by BVI, and grey curves in (d) and (e) show the component distributions obtained using SMVI. In the bottom two panels, the Y-axis range is changed to highlight the individual component distributions.}
	\label{fig:1D_mixture_Gaussian}
\end{figure}

We use a simple 1-dimensional (1D) toy example to compare the above methods. The target distribution is a mixture of two Gaussian distributions $p(x) = 0.6\mathcal{N}(x;-1, 0.5) + 0.4 \mathcal{N}(x;1, 0.7)$, shown by a red curve in Figure \ref{fig:1D_mixture_Gaussian}. Blue histograms in panels a, b, c, d and e in Figure \ref{fig:1D_mixture_Gaussian} show variational distributions obtained using ADVI, SVGD with 500 particles, BVI with 20 components, SMVI with 2 components, and SMVI with 20 components, respectively. Since ADVI uses only a single Gaussian distribution, the approximation fails to capture the two modes correctly. In Figure \ref{fig:1D_mixture_Gaussian}b, grey vertical lines show the locations of the optimised particles, which are distributed around the two modes, and the corresponding blue histogram approximates the target pdf. However, the SVGD approximation is represented by a finite set of particles (point estimates) rather than by a continuous distribution, making the representation less efficient. Figure \ref{fig:1D_mixture_Gaussian}c shows the results obtained using BVI. Each component is optimised sequentially, and the first component, shown by a grey curve, recovers a unimodal solution that is similar to that obtained using ADVI. By adding more components sequentially, the solution starts to capture the bimodal structure correctly. The results with 20 components recover the target distribution reasonably accurate. However for the BVI configuration used here, it is almost impossible to provide an accurate approximation to this target distribution using only two components, because the first component is fixed after it has been optimised. As we show in a full waveform inversion example below in Section \ref{section:fwi}, this property makes the method less efficient.

In Figures \ref{fig:1D_mixture_Gaussian}d and e, grey curves show the component distributions obtained using SMVI. In panel d, SMVI provides a reasonably accurate result with only 2 components, since the target distribution is itself a mixture of 2 Gaussians. The left Gaussian component is biased towards the left, because the target distribution uses unequal mixture weights (0.6 and 0.4), whereas the SMVI components are assigned equal weights according to equation \ref{eq:mixture}. The inaccuracy of the \textit{a priori} weight coefficients leads to biased Gaussian mean estimation. Nevertheless, the optimised mixture approximation still recovers the target distribution reasonably accurately, because the means and variances of the 2 components are adjusted during optimisation to fit the target distribution. This suggests that, at least in simple cases, equal weight mixtures can provide sufficient flexibility to represent an asymmetric multimodal target distribution. In addition, each grey curve shows that the corresponding component provides information about parameter space with non-zero width, so the mixture distribution gives a continuous representation of the target distribution. Given that we never know the true posterior solution in practical applications, we never know the optimal number of components to be used \textit{a priori}, so Figure \ref{fig:1D_mixture_Gaussian}e shows SMVI results obtained using 20 components, which also recovers the two modes. This approximation looks worse than with 2 components, since we use 20 components with equal weights to approximate a distribution with 2 components. Therefore, 18 redundant components are assigned a weight coefficient of 0.05 rather than 0, which introduces additional error to the final mixture distribution. Still, despite the additional components, the final result still represents the correct two modes and shows similar accuracy compared to that obtained using BVI with 20 components. In the bottom two panels of Figure \ref{fig:1D_mixture_Gaussian}, the Y-axis range is changed to display the individual component distributions fully.

Table \ref{table:comparison} lists some key properties of these methods. By building a mixture variational distribution, SMVI combines advantages of both ADVI and SVGD. In Section \ref{section:example}, several examples are used to demonstrate these properties.

\begin{table}
	\caption{Comparison of key properties among ADVI, PSVI, SVGD, BVI and SMVI. Computational cost is compared and discussed in Table 2 below. Note that in the toy example, the computational cost of BVI and SMVI is similar, and in a FWI example presented in Section \ref{section:fwi}, the cost of BVI is higher than that of SMVI. Therefore, we describe the relative computational cost of BVI as intermediate to higher.}
	\centering
	\begin{tabular}{lllll}
		\hline
		Property & ADVI and PSVI & SVGD & BVI & SMVI \\
		\hline
		Representation type & Parametric & Particle-based & Parametric mixture & Parametric mixture \\
		Posterior flexibility & Limited & Flexible & Flexible & Flexible \\
		Multimodality representation & Unimodal & Multimodal & Multimodal & Multimodal \\
		Generation of new samples  & Direct & {Not direct} & {Direct} & {Direct} \\
		Relative computational cost & {Lower} & {Higher} & {Intermediate to Higher} & {Intermediate} \\
		\hline
	\end{tabular}
	\label{table:comparison}
\end{table}

\section{Examples}
\label{section:example}
In this section, we apply the above suite of methods to two synthetic seismic inverse problems to test the effectiveness of SMVI, and to prove different properties listed in Table \ref{table:comparison}. Specifically in a surface wave inversion example, we prove that SMVI can provide accurate inversion results with multi-modal posterior solutions; in seismic full waveform inversion (FWI), we demonstrate that SMVI is computationally efficient and is able to mitigate mode collapse that appears in the SVGD solution. Finally we apply the method to perform Love wave tomography of the British Isles in a real data application. Since SMVI bridges ADVI (PSVI) and SVGD, we compare these three types of methods in all presented examples in order to identify the advantages of SMVI. In the FWI example, we also compare the performance of BVI which serves as an independent method in a truly complex and difficult case, in which a Monte Carlo solution is unavailable.

\subsection{Surface wave inversion}
\begin{figure}
\centering\includegraphics[width=\textwidth]{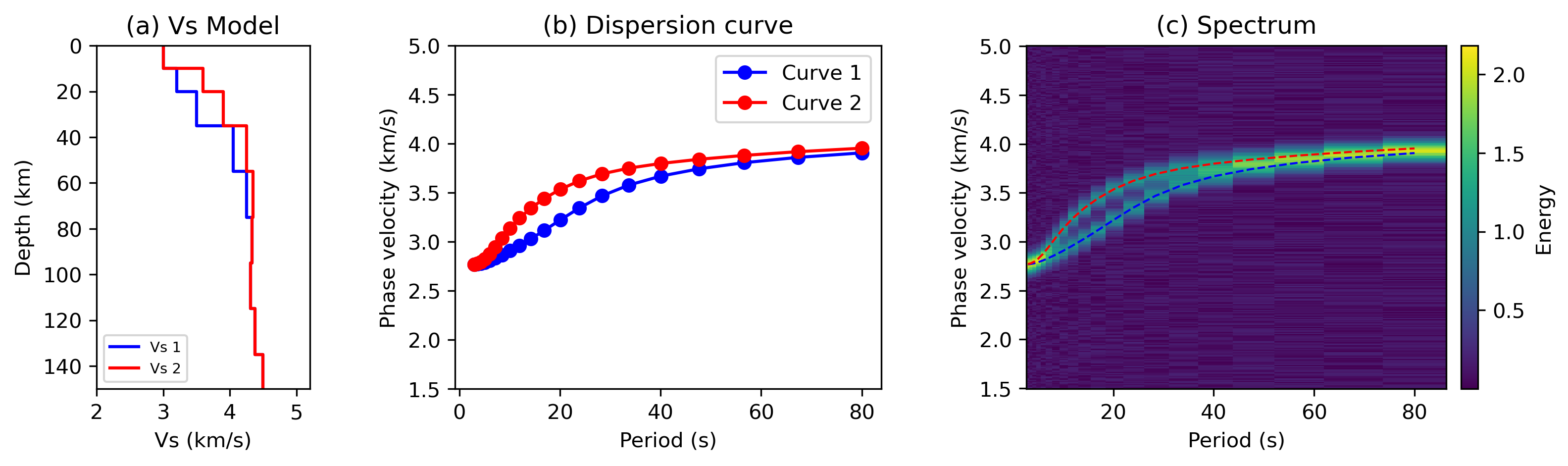}
\caption{(a) Two shear wave velocity ($v_s$) profiles used in the surface wave inversion example. Shear wave velocity values in layers 2--5 are lower in the blue profile (denoted as $v_s$ 1) than in the red profile ($v_s$ 2). (b) Two fundamental mode Rayleigh wave phase velocity dispersion curves corresponding to the two S-wave profiles in (a). (c) A synthetic energy spectrum constructed from the two dispersion curves in (b). Maximum energy is assigned along each dispersion curve and decreases with increasing distance from the curves. Dashed blue and red curves correspond to the two dispersion curves in (b).}
\label{fig:swi_model_data_spectrum}
\end{figure}

In surface wave inversion, spatially 1D depth-dependent subsurface shear wave (S-wave) velocity structures are estimated using recorded surface wave dispersion data. We design a surface wave inverse problem with a multimodal Bayesian posterior solution. Specifically, we define two vertical 9-layer S-wave velocity profiles, represented by blue and red lines in Figure \ref{fig:swi_model_data_spectrum}a. Shear wave velocity values in layers 2--5 of the blue profile (denoted as profile 1) are lower than those of the red profile (profile 2). We then perform forward simulations and predict the corresponding fundamental mode Rayleigh wave phase velocity dispersion curves using the SURF96 program \citep{herrmann2013computer}, as displayed by blue and red curves in Figure \ref{fig:swi_model_data_spectrum}b, respectively. Given these two dispersion curves, we construct a spectrum in which the amplitude (energy) is peaked along the modelled dispersion curves and decays with increasing distance from the curves, as shown in Figure \ref{fig:swi_model_data_spectrum}c. This spectrum has two clearly separated branches of energy, which could arise in practice from multipathing of the surface waves through a laterally heterogeneous medium consisting of the two neighbouring structures shown in Figure \ref{fig:swi_model_data_spectrum}a \citep{zhang2023surface}. In real applications, such a spectrum can be obtained by performing frequency-phase velocity analysis of recorded waveform data, and in this example we treat this spectrum as an observed dataset and invert it to estimate possible S-wave velocity profiles.

In conventional surface wave inversion, a dispersion curve must first be picked from the spectrum in Figure \ref{fig:swi_model_data_spectrum}c, by selecting a surface wave velocity value with the highest energy at each frequency. Either the dashed blue or red dispersion curve in Figure \ref{fig:swi_model_data_spectrum}c can be picked. Subsequently, a dispersion inversion is performed to find S-wave models that fit the selected dispersion curve to within its uncertainty. However, the inversion results then fit only one branch of energy in the spectrum and can not describe the multimodality of the surface wave data accurately. To address this problem, \cite{zhang2023surface} introduced a Bayesian surface wave inversion method using an energy likelihood function, in which an energy spectrum such as that in Figure \ref{fig:swi_model_data_spectrum}c is used directly as an observed dataset for inversion. The inversion results thus include shear wave velocity profiles that fit either branch of energy and correctly include multimodality that may occur in dispersion data. More details of the energy likelihood function method can be found in \cite{zhang2023surface}, and we use this likelihood function to invert the spectrum displayed in Figure \ref{fig:swi_model_data_spectrum}c.

\begin{figure}
	\centering\includegraphics[width=\textwidth]{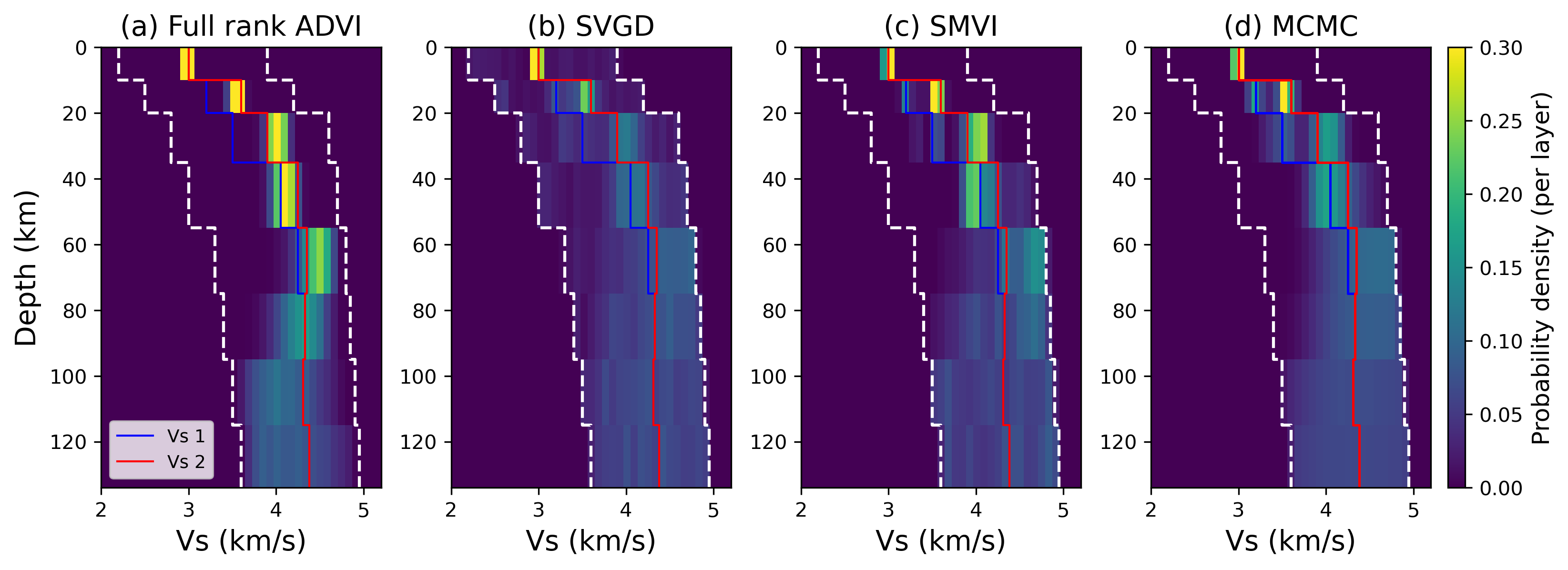}
	\caption{Posterior marginal distributions of shear wave velocity values within each layer, obtained using (a) full rank ADVI, (b) SVGD, (c) SMVI, and (d) McMC. Solid blue and red lines denote the two reference S-wave velocity profiles displayed in Figure \ref{fig:swi_model_data_spectrum}a, and white dashed lines represent the uniform prior bounds for each layer.}
	\label{fig:swi_posterior_marginals}
\end{figure}

For the purposes of this purely illustrative inversion, we set layer thicknesses to their true values and estimate only S-wave velocity values ($v_s$) in different layers, resulting in a 9-dimensional inverse problem. P-wave velocity ($v_p$) and density ($\rho$) are calculated from shear wave velocity using $v_p = 1.77 v_s$ and $\rho = 0.32v_p + 0.77$ \citep{berteussen1977moho}. Similar formulae were applied in the forward problem used to construct dispersion curves in Figure \ref{fig:swi_model_data_spectrum}b. The only reason that we avoid committing an inverse crime (in which synthetic data are constructed from exactly the same forward problem as used in the inversion) is by adding the finite width of energy around each dispersion curve in Figure \ref{fig:swi_model_data_spectrum}c. A uniform prior distribution is used in this example, and its lower and upper bounds are shown for the parameter in each depth range by two dashed white lines in Figure \ref{fig:swi_posterior_marginals}. We apply full rank ADVI, SVGD and SMVI to this problem. We also perform a Metropolis-Hastings Markov chain Monte Carlo (MH-McMC) inversion for comparison. Because both the dimensionality and the computational cost of forward simulations are relatively low, each algorithm is run for a sufficiently long time to achieve apparent convergence of the results shown, allowing a fair comparison among different methods. For full rank ADVI, we optimise a single transformed Gaussian distribution with a full covariance matrix to approximate the posterior pdf. For SVGD, we optimise 1,000 particles for 40,000 iterations (similar results are obtained for 1,500 particles and results deteriorate for 500 particles, so we assume that a choice of 1,000 particles is sufficient to represent performance). For SMVI, a mixture variational distribution with 50 components is used to approximate the posterior solution, and each component is parametrised by a transformed uncorrelated Gaussian distribution, similarly to mean field ADVI. For McMC, 40 chains are run in parallel, each drawing 1 million samples, and solutions did not visibly improve using either more chains or more samples per chain.

Figure \ref{fig:swi_posterior_marginals} displays the posterior marginal distributions of shear wave velocity values obtained using the 4 tested methods. The McMC results (Figure \ref{fig:swi_posterior_marginals}d) serve as a reference Bayesian posterior solution for this problem. Two distinct modes are visible in the second and third layers, reflecting the multimodal nature of the posterior pdf induced by the synthetic surface wave spectrum in Figure \ref{fig:swi_model_data_spectrum}c. In addition, the two reference $v_s$ profiles (blue and red lines) each lie within one of the two high probability regions, indicating that this synthetic example indeed produces a multimodal posterior solution and that McMC recovers these two modes correctly.

Results from full rank ADVI (Figure \ref{fig:swi_posterior_marginals}a) differ substantially from the McMC solution. The posterior approximation concentrates around a single mode and fails to recover the second high probability mode identified by McMC. This is expected because the method uses a single transformed Gaussian distribution and therefore can not represent a multimodal posterior pdf adequately. Note that the estimated posterior distribution does not span both modes as it did in Figure \ref{fig:1D_mixture_Gaussian}a. The latter behaviour occurs if modes are fairly close together; in this surface wave problem the modes are well separated as shown in Figure \ref{fig:swi_posterior_marginals}d, so ADVI converges to one or other mode. In Figure \ref{fig:swi_posterior_marginals}b, the SVGD results partially capture the multimodal structure by recovering two modes in the marginal pdf's in the second and third layers. However, compared with the results from McMC, the posterior approximation remains considerably more diffuse. In particular, noticeable probability density is distributed near both the lower and upper prior bounds in several layers, suggesting that a substantial fraction of particles fails to move into the principal high probability regions. This indicates that the algorithm fails to fully converge to the correct, target posterior distribution, leading to a less accurate approximation. In contrast, results from SMVI (Figure \ref{fig:swi_posterior_marginals}c) recover the principal characteristics of the McMC solution. The bimodal distributions in the second and third layers are captured accurately, and the posterior distributions in the remaining layers closely match those obtained from McMC. Overall, the agreement between the SMVI and McMC results demonstrates that SMVI can approximate complex multimodal posterior distributions accurately while avoiding the single mode limitation of ADVI, at least in this 9-dimensional problem.

\begin{figure}
\centering\includegraphics[width=\textwidth]{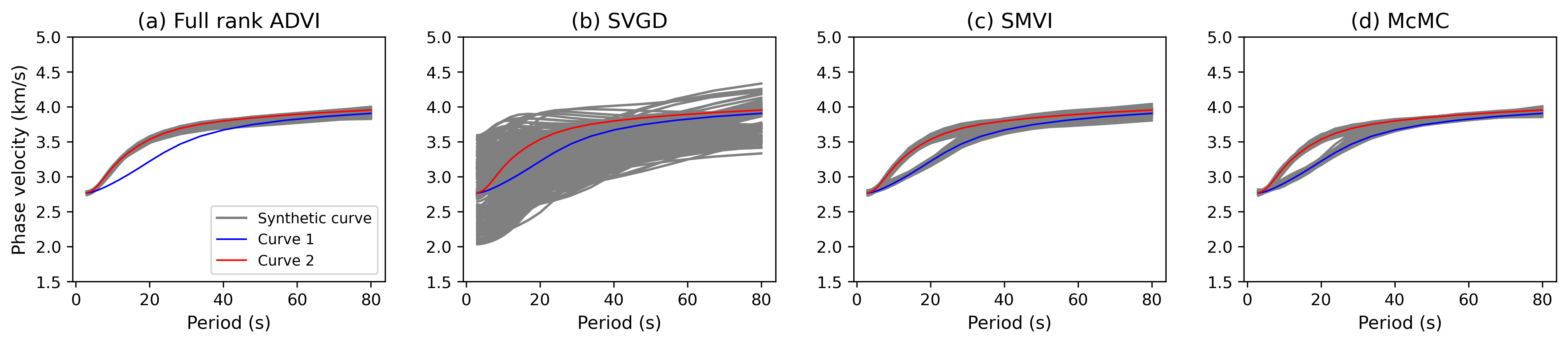}
\caption{Synthetic phase velocity dispersion curves (grey lines) simulated using posterior samples from (a) full rank ADVI, (b) SVGD, (c) SMVI, and (d) McMC. In each panel, the red and blue lines represent the two reference (observed) dispersion curves shown in Figure \ref{fig:swi_model_data_spectrum}b.}
\label{fig:swi_data_fit}
\end{figure}

To further evaluate the accuracy of the posterior distributions obtained using the 4 methods, we draw random samples from each posterior pdf and perform forward simulations to generate the corresponding dispersion curves. Note that for SVGD, we simply use the obtained 1,000 particles. The results are shown in Figure \ref{fig:swi_data_fit}. Grey curves represent simulated dispersion curves, and blue and red curves correspond to the two reference dispersion curves shown in Figure \ref{fig:swi_model_data_spectrum}b.

Synthetic data generated from posterior samples obtained using full rank ADVI are concentrated around the red dispersion curve. This behaviour is consistent with the posterior distribution shown in Figure \ref{fig:swi_posterior_marginals}a, where full rank ADVI captures only a single posterior mode around the red $v_s$ profile 2. This again demonstrates that a single transformed Gaussian approximation can not describe the multimodality of surface wave data accurately. The SVGD results exhibit substantially larger variability, and many of the simulated dispersion curves do not coincide closely with either of the two reference curves. This is consistent with the posterior distribution in Figure \ref{fig:swi_posterior_marginals}b, in which a considerable number of particles remain distributed outside of the principal high probability regions. Consequently, this supports the interpretation above that the algorithm may converge to an inaccurate approximation of the Bayesian posterior solution. In contrast, synthetic dispersion curves generated from SMVI closely resemble those obtained from McMC. In both cases, the simulated data cluster around the two reference dispersion curves, indicating that SMVI approximates the multimodal posterior distribution accurately and produces inversion results comparable to those obtained using McMC.

\subsection{Full Waveform Inversion}
\label{section:fwi}

\begin{figure}
	\centering\includegraphics[width=\textwidth]{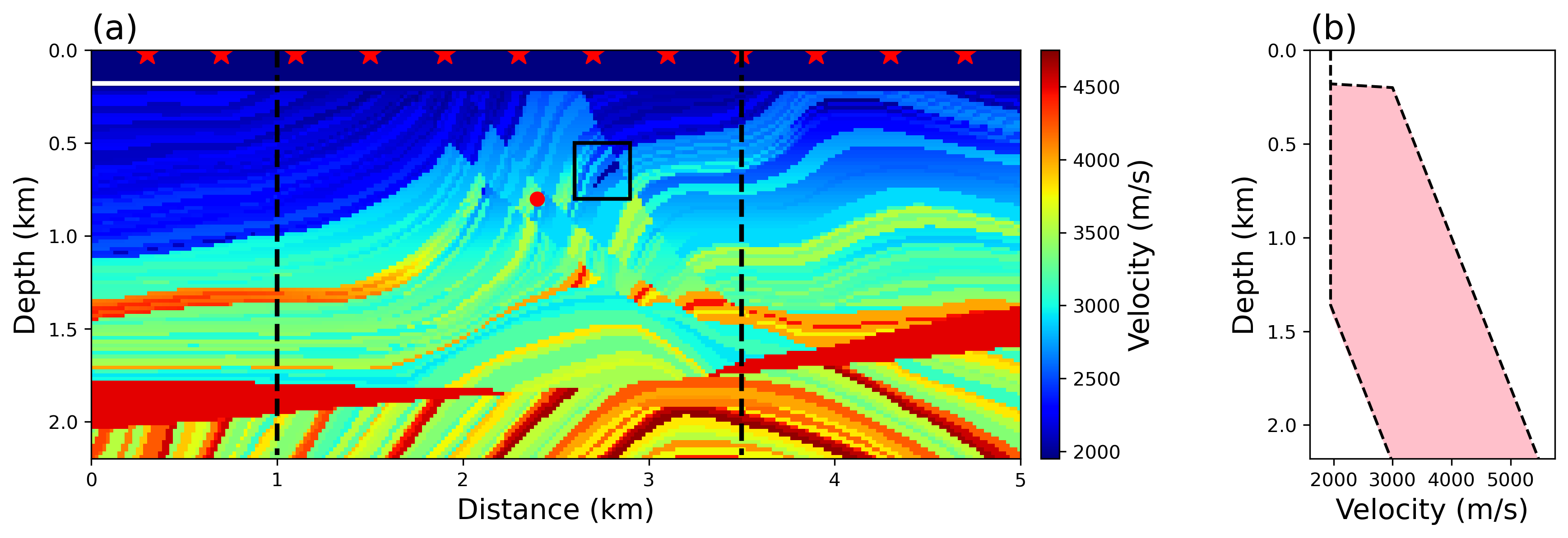}
	\caption{(a) True Marmousi velocity model used in the acoustic FWI example. Source locations are indicated by red stars and a receiver line by a white line. Vertical dashed black lines indicate locations of two profiles used to compare inversion results in the main text. (b) Uniform prior bounds at different depths, represented by two dashed black lines.}
	\label{fig:fwi_vel_true_prior}
\end{figure}

In the second example, the performance of SMVI is tested using a spatially 2D acoustic full waveform inversion example. As displayed in Figure \ref{fig:fwi_vel_true_prior}a, the true velocity model is a truncated Marmousi model \citep{martin2006marmousi2}, discretised into 110 $\times$ 250 regular grid cells with a cell size of 20 m in both directions. A simple acquisition geometry is used, in which 12 sources (red stars in Figure \ref{fig:fwi_vel_true_prior}a) and 250 receivers (distributed evenly along the horizontal white line) are deployed at 20 m and 200 m depth, respectively. A Ricker wavelet with a dominant frequency of 10 Hz is used to generate a waveform dataset using a time domain finite difference method. Gaussian random noise with zero mean and a standard deviation equal to 1\% of the average maximum amplitude is added to the synthetic data, and the resulting noisy data are used as observed waveform data in this test. A uniform prior distribution is defined for velocity values in each grid cell, and the lower and upper prior bounds at different depths are represented by dashed black lines in Figure \ref{fig:fwi_vel_true_prior}b. Velocity values in the water layer (down to 200 m depth) are fixed at their true values during inversion. The likelihood function is defined as an uncorrelated Gaussian distribution, with diagonal values equal to the magnitude of the random noise added to the noise-free waveform data, to mimic a situation in which the data noise level is estimated accurately. The gradient of the log-likelihood function with respect to a given velocity model is computed using the adjoint-state method \citep{plessix2006review}.

\subsubsection{\textbf{Inversion results from different methods}}
We first apply mean field ADVI, PSVI, SVGD and SMVI to this FWI problem, and compare the resulting posterior approximations to demonstrate the advantages of SMVI obtained by combining ADVI and SVGD. Given the dimensionality of this problem (25,000), our McMC methods failed to converge, and full rank ADVI is intractable in terms of both memory requirements and computational cost. Both mean field ADVI and PSVI employ a transformed Gaussian distribution to approximate the posterior solution, with the former assuming a diagonal covariance matrix and the latter using a structured sparse covariance matrix that models known, dominant inter-parameter correlations. Following \cite{zhao2024physically}, we use a 5$\times$5 correlation kernel for PSVI, meaning that for each model parameter, correlations are modelled with parameters in the surrounding local neighbourhood extending up to two grid cells away in both spatial directions. For both algorithms, 5,000 iterations are performed for variational optimisation, with 2 samples per iteration used to approximate ELBO and its gradient (equation \ref{eq:grad_elbo}). For SVGD, 500 particles are optimised for 400 iterations. For SMVI, 100 components are used to construct a mixture distribution, and are optimised for 400 iterations. Within each iteration, 1 sample is used to estimate ELBO and its gradient (equation \ref{eq:grad_elbo}). 

\begin{figure}
	\centering\includegraphics[width=\textwidth]{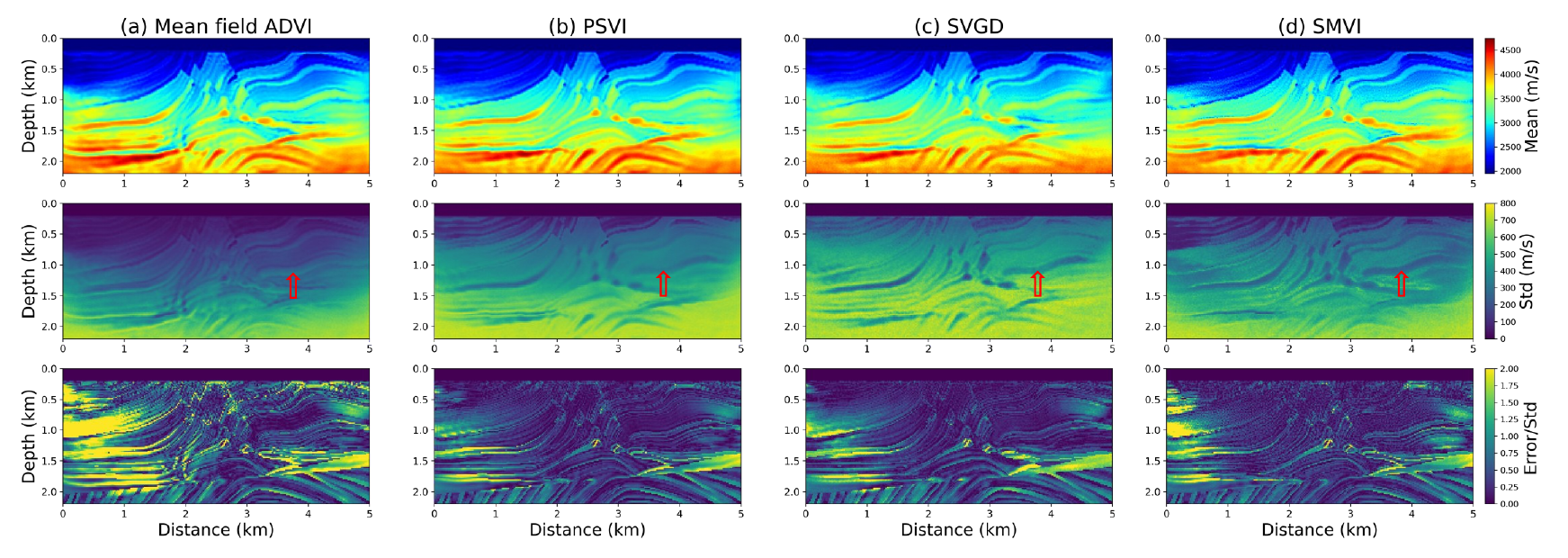}
	\caption{Mean (top row), standard deviation (middle row), and relative error (bottom row) of the approximate posterior distributions obtained using (a) mean field ADVI, (b) PSVI, (c) SVGD, and (d) SMVI, respectively. The relative error is calculated by dividing the absolute difference between the posterior mean and the true velocity value by the posterior standard deviation at each point.}
	\label{fig:fwi_4methods_mean_std_error}
\end{figure}

Figure \ref{fig:fwi_4methods_mean_std_error} compares several posterior statistics obtained using the 4 methods. The first row shows the posterior mean velocity models, the second row presents the posterior standard deviation (uncertainty) maps, and the third row displays the corresponding relative error maps, computed for each cell individually by dividing the absolute error between the true and mean models by the standard deviation model. Overall, the posterior mean velocity models obtained using the 4 methods reproduce the principal geological structures and are all broadly consistent with the true velocity model (Figure \ref{fig:fwi_vel_true_prior}a). Main velocity interfaces and velocity anomalies in the central part of the model are reconstructed accurately in all 4 cases, indicating that the dominant velocity features are resolved reasonably well by all methods.

Clearer differences are observed in the posterior standard deviation maps. Compared with PSVI, SVGD and SMVI, mean field ADVI yields substantially different uncertainty estimates in several regions. For example, around a red arrow, mean field ADVI predicts noticeably higher posterior standard deviation values than the surrounding regions, whereas the other three methods yield relatively lower uncertainty values. Similar discrepancies are observed near several major velocity interfaces. These results are consistent with previous studies \cite[e.g.,][]{zhang20233, zhao2024physically}, which showed that mean field ADVI tends to provide biased uncertainty estimates because its factorised Gaussian assumption neglects correlations among model parameters. In contrast, posterior standard deviation maps obtained using PSVI, SVGD and SMVI are highly consistent, with similar spatial distributions and uncertainty levels throughout the model. The close agreement among these three methods suggests that they infer similar uncertainty structures, at least at the level of these second-order statistics. The relative error maps in the third row further support these observations: mean field ADVI exhibits larger reconstruction errors in several regions, whereas PSVI, SVGD and SMVI give consistent spatial distributions with generally lower relative errors (mostly smaller than 2), indicating more reliable uncertainty estimates.

\begin{figure}
	\centering\includegraphics[width=\textwidth]{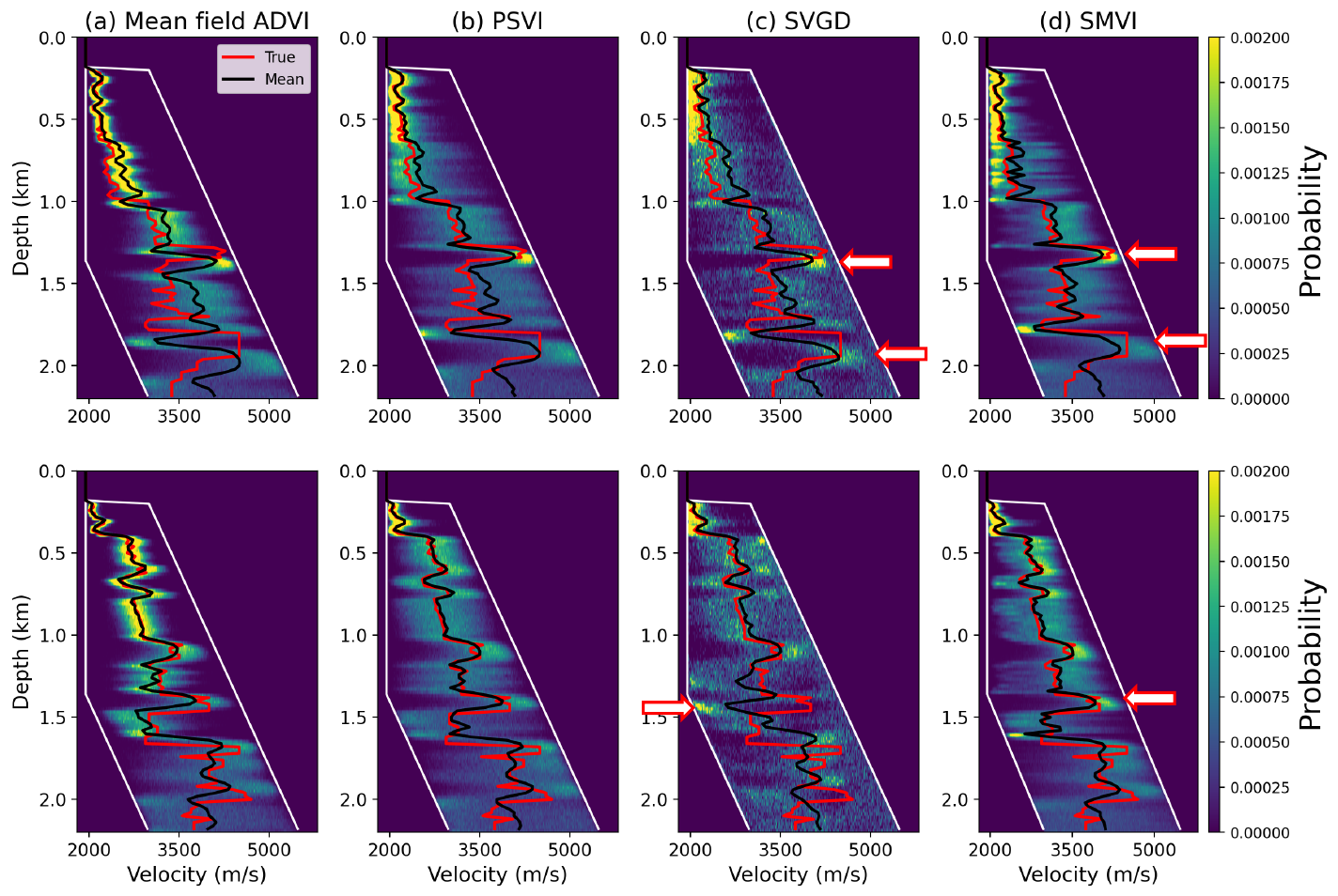}
	\caption{Posterior marginal distributions along two vertical velocity profiles at horizontal locations of 1 km (top row) and 3.5 km (bottom row), whose locations are marked by black dashed lines in Figure \ref{fig:fwi_vel_true_prior}a. From left to right, the columns show results obtained using (a) mean field ADVI, (b) PSVI, (c) SVGD, and (d) SMVI, respectively.}
	\label{fig:fwi_4methods_2marginals}
\end{figure}

Figure \ref{fig:fwi_4methods_2marginals} further compares posterior marginal distributions along two vertical velocity profiles located at horizontal locations of 1 km and 3.5 km, indicated by vertical dashed black lines in Figure \ref{fig:fwi_vel_true_prior}a. Red and black curves in each panel represent the true velocity and posterior mean profiles, respectively. Consistent with the posterior standard deviation maps in Figure \ref{fig:fwi_4methods_mean_std_error}, the marginal pdf's obtained from mean field ADVI are generally narrower than those from the other methods, indicating underestimated posterior uncertainty due to the factorised Gaussian assumption. In contrast, PSVI, SVGD and SMVI yield broadly similar marginal distributions, further suggesting that these methods capture the main uncertainty structure of the posterior solution.

However, localised differences between the SVGD and SMVI results are displayed, which are not apparent from the uncertainty maps in Figure \ref{fig:fwi_4methods_mean_std_error}. In the top row of Figure \ref{fig:fwi_4methods_2marginals}, highlighted by white arrows, high probability regions in the results from SVGD are shifted from the true depths of the high velocity layers, possibly because of cycle skipping (i.e., waveform phase misalignment between forward modelled and observed waveforms exceeding half of a sinusoidal cycle leads the inversion towards an incorrect solution in which it attempts to fit the adjacent cycle instead). The corresponding high probability regions in the SMVI posterior solution are centred around the correct locations. A similar behaviour is observed in the bottom row around the white arrow where the true velocity profile contains a high velocity layer. Results from SVGD show most probability density mass around a lower velocity zone, whereas the SMVI posterior solution displays high probability close to the correct high velocity structure. These results suggest that the posterior approximation from SMVI is more accurate than that from SVGD in this example, despite their similar global posterior statistics displayed in Figure \ref{fig:fwi_4methods_mean_std_error}.

A possible explanation for this difference in performance is the particle degeneracy reported previously for SVGD in high dimensional inverse problems \citep{ba2021understanding, zhang20233}. In this example, 500 particles are used to approximate a posterior pdf in a 25,000-dimensional parameter space, and of course a storeable finite set of particles (point estimates) can not represent such a high dimensional target distribution adequately. In contrast, by assigning each particle a neighbourhood in parameter space, each SMVI component explores continuous parameter space through a distribution with finite width rather than through a single point estimate. This property may contribute to mitigation of the mode collapse problem in SVGD.

\begin{figure}
	\centering\includegraphics[width=\textwidth]{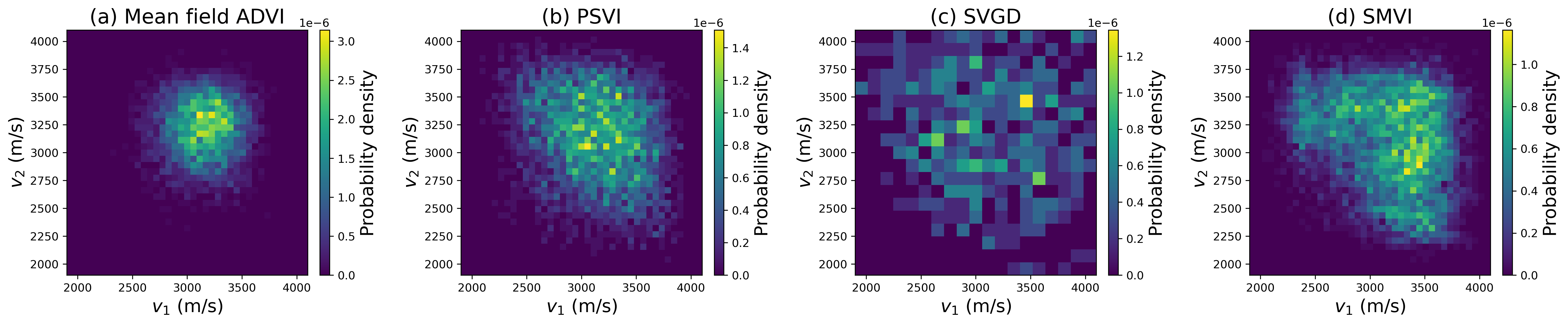}
	\caption{Posterior bivariate marginal distributions of velocity values at two vertically adjacent locations, (2.4 km, 0.84 km) and (2.4 km, 0.86 km), obtained from different inversion methods. Their locations are marked by a red circle in Figure \ref{fig:fwi_vel_true_prior}a.}
	\label{fig:fwi_4methods_bi_2dmarginals}
\end{figure}

Figure \ref{fig:fwi_4methods_bi_2dmarginals} shows posterior bivariate marginal distributions of velocity values at two vertically adjacent locations, (2.4 km, 0.84 km) and (2.4 km, 0.86 km). Their locations are marked by a red circle in Figure \ref{fig:fwi_vel_true_prior}a. Again, mean field ADVI provides a narrow posterior distribution, and both ADVI and PSVI show Gaussian-shaped bivariate distributions because of the transformed Gaussian variational assumption. In contrast, results from SVGD and SMVI show clear non-Gaussian features, demonstrating that these methods can represent more complex posterior solutions. Note that the distribution from SVGD in Figure \ref{fig:fwi_4methods_bi_2dmarginals}c remains sparse even with large pixels used for illustration, since only 500 particles are used to represent the posterior pdf in this 2D space (around 23 samples per dimension).

\begin{figure}
	\centering\includegraphics[width=\textwidth]{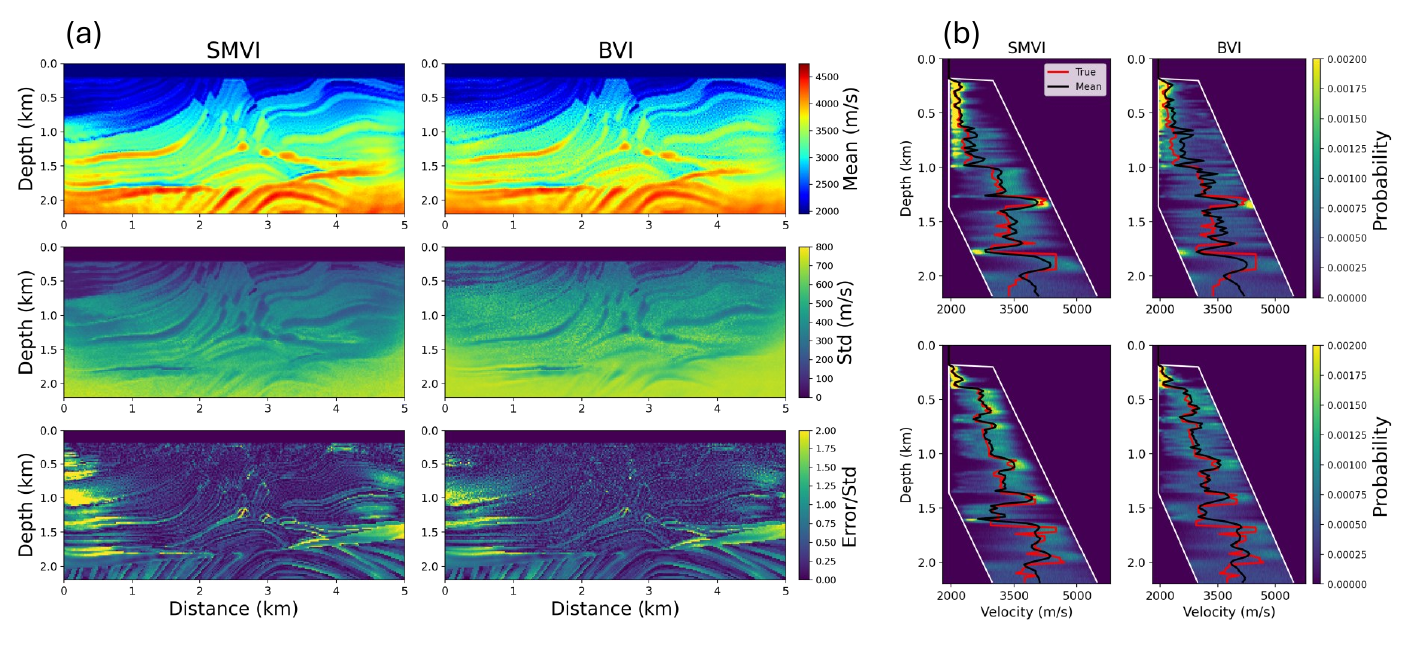}
	\caption{(a) Posterior mean (top row), standard deviation (middle row) and relative error (bottom row) maps obtained using SMVI and BVI, similarly to those displayed in Figure \ref{fig:fwi_4methods_mean_std_error}. (b) Posterior marginal distributions along two vertical profiles from SMVI and BVI, similarly to those displayed in Figure \ref{fig:fwi_4methods_2marginals}.}
	\label{fig:fwi_smvi_bvi}
\end{figure}

Given that SMVI combines ADVI and SVGD and is thus not completely independent of those two methods, we apply BVI to this FWI problem, which serves as an independent method with which to cross-validate the results from SMVI. In addition, since both SMVI and BVI employ a mixture variational distribution, the main difference being in the way that the mixture distribution is constructed, it is necessary to compare their performance in terms of inversion accuracy and efficiency. For BVI, 24 uncorrelated transformed Gaussian components are optimised sequentially, and each component is updated for 2500 iterations using 2 samples per iteration to estimate ELBO. In principle any new BVI component is introduced to improve the accuracy of the existing mixture distribution. We observe that the posterior statistics stop changing significantly after optimising 24 components, meaning that the approximate posterior solution might be reasonably good at this point. As shown in Figure \ref{fig:fwi_smvi_bvi}, the results from BVI and SMVI are generally consistent, and both are similar to those obtained using PSVI and SVGD displayed in Figures \ref{fig:fwi_4methods_mean_std_error} and \ref{fig:fwi_4methods_2marginals}, meaning that all four algorithms are able to find reasonably accurate Bayesian posterior statistics. 

Table \ref{table:fwi_cost} compares the computational cost and optimisation complexity of the 5 tested methods. The computational cost is measured by the total number of forward and adjoint simulations, since they are by far the most expensive part of each algorithm. The optimisation complexity is characterised by the total number of parameters optimised in each inversion. For this acoustic FWI example, the true velocity model contains 25,000 unknown parameters. Consequently, 50,000 parameters are optimised in mean field ADVI, defining the mean and standard deviation of a single uncorrelated Gaussian distribution. In PSVI, additional covariance parameters are introduced to capture dominant parameter correlations, increasing the optimisation problem to 1.55 million parameters. SVGD employs 500 model particles, resulting in 12.5 million optimisation variables. In SMVI, 100 particles are used to parametrise 100 uncorrelated Gaussian components, containing 5 million parameters in total. In BVI, 24 uncorrelated Gaussian components, and thus 1.2 million parameters in total, are optimised.

\begin{table}
	\caption{Comparison of computational cost and optimisation complexity of the 5 methods tested in the FWI example. The computational cost is quantified by the total number of forward or adjoint simulations, and the optimisation complexity by the total number of parameters optimised during inversion. Note that the true model contains 25,000 unknown velocity parameters. In the fourth column, the numbers per iteration represent the number of samples used to estimate the ELBO and its gradient for mean field ADVI, PSVI, BVI and SMVI (equation \ref{eq:grad_elbo}). For SVGD, one simulation is performed per iteration per particle. In the fifth column, the total number of simulations for each method is obtained by multiplying the numbers in the second, third, and fourth columns. Wall clock time in the sixth column is recorded based on an Intel Xeon Max 9480 processor.}
	\centering
	\begin{tabular}{ccccccc}
		\hline
		Method & \makecell{Component(s) \\ or particles} &
		\makecell{Iterations} &
		\makecell{Simulations \\ per iteration} &
		\makecell{Total \\simulations} &
		\makecell{Wall clock \\ time (hours)} &
		\makecell{Total parameters \\ optimised} \\
		\hline 
		MF ADVI & 1 & 5,000 & 2 & 10,000 & 25 & 50,000 \\
		PSVI & 1 & 5,000 & 2 & 10,000 & 26 & 1,550,000 \\
		BVI & 24 & 2,500 & 2 & 120,000 & 260 & 1,200,000 \\
		SVGD & 500 & 400 & 1 & 200,000 & 68 & 12,500,000 \\
		SMVI & 100 & 400 & 1 & 40,000 & 14 & 5,000,000 \\
		\hline
	\end{tabular}
	\label{table:fwi_cost}
\end{table}

The optimisation complexity of SVGD is therefore substantially higher than that of the other methods. This more difficult optimisation problem may contribute to the less accurate inversion results observed in both this example and the surface wave inversion example (Figures \ref{fig:swi_posterior_marginals} and \ref{fig:swi_data_fit}). By contrast, SMVI requires fewer than half the number of optimisation variables while providing a richer continuous representation of the posterior pdf through a Gaussian mixture distribution. This mitigates the particle degeneracy problem commonly observed in SVGD and yields more accurate posterior approximations with lower optimisation complexity.

From the perspective of computational cost, mean field ADVI and PSVI require the fewest simulations. While previous studies have shown that the latter provides better results than the former \cite[e.g.,][]{zhao2024physically}, both remain restricted to unimodal Gaussian posterior approximations and are therefore inaccurate for strongly multimodal or highly non-Gaussian inverse problems. SVGD, BVI and SMVI can represent multimodal problems, and of these SVGD requires the highest number of simulations. In BVI, 24 components are optimised whereas in SMVI 100 components are optimised, which seems to suggest that BVI is more efficient than SMVI. However, 120,000 forward simulations are performed in BVI, 3 times more than were performed in SMVI. This is because in BVI hyper-parameters of each component are optimised directly using a gradient-based optimisation. In SMVI, those hyper-parameters are optimised in a reproducing kernel Hilbert space, in which the algorithm converges faster with fewer iterations \citep{liu2016kernelized}. In addition, as mentioned above, the 24 BVI components are optimised sequentially, resulting in a wall clock time of about 260 hours using an Intel Xeon Max 9480 processor; in contrast, 100 SMVI components are optimised simultaneously, which takes about 14 hours using the same hardware, since the algorithm makes full use of all available HPC resources. As listed in the sixth column in Table \ref{table:fwi_cost}, SMVI takes the shortest wall clock time, which also demonstrates the practical efficiency of SMVI over BVI (and other methods) for large scale inverse problems.

Beyond the computational cost quantified above, an additional practical distinction between BVI and SMVI concerns the effort required to identify suitable optimisation settings. The comparison in Table \ref{table:fwi_cost} reports the cost after a relatively optimal configuration has been selected for each method, but does not include the problem-specific tuning required to obtain such a configuration. Across our experience of applying BVI and SMVI to various geophysical problems, including the FWI example presented herein, SMVI has generally produced useful posterior approximations with relatively little adjustment of its optimisation settings. By contrast, BVI typically required exploration of several hyper-parameter configurations before comparable results were obtained. This difference is likely caused by the distinct optimisation mechanisms used by the two methods. In BVI, component parameters are optimised directly and sequentially using stochastic gradient estimates, whereas SMVI jointly updates the component ensemble within a reproducing kernel Hilbert space which may provide a less sensitive optimisation procedure in practice \citep{liu2016kernelized}. Our experience thus suggests that SMVI can impose a lower problem-specific tuning burden than BVI, potentially further reducing the end-to-end computational effort required when applying the method to new inversion problems.


\subsubsection{\textbf{Detailed comparison between SVGD and SMVI}}
\begin{figure}
	\centering\includegraphics[width=\textwidth]{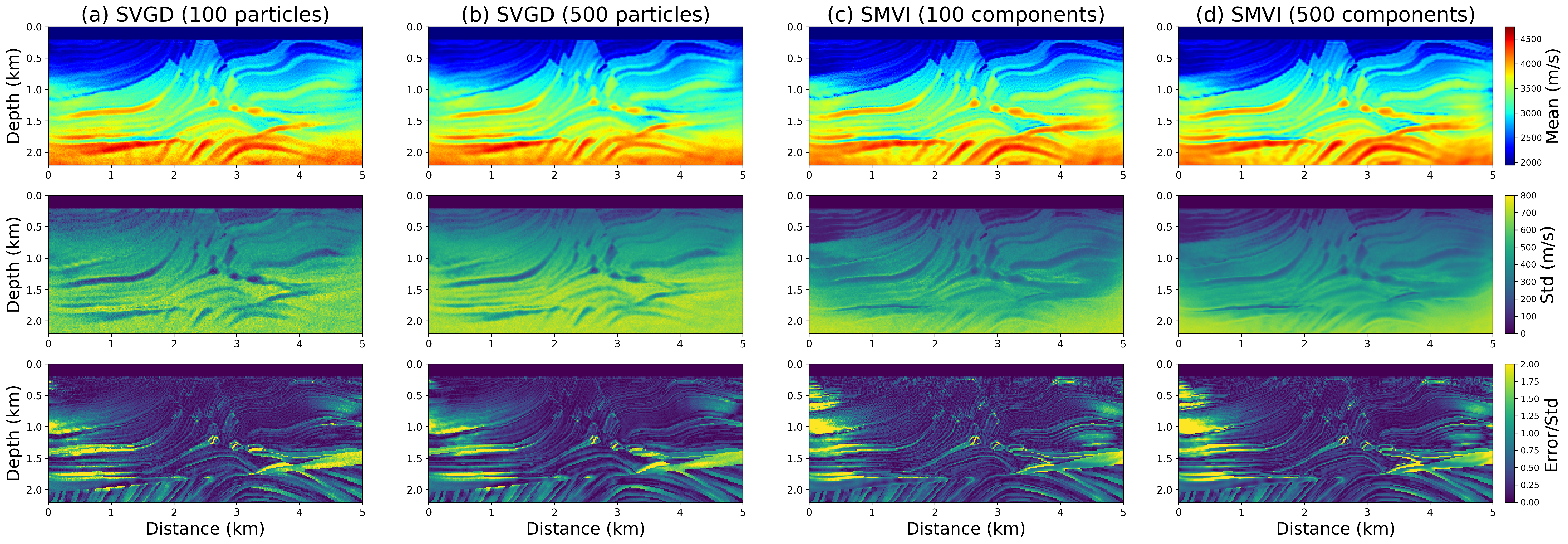}
	\caption{Mean (top row), standard deviation (middle row), and relative error (bottom row) maps of approximate posterior distributions obtained using (a) and (b) SVGD with 100 and 500 particles, and (c) and (d) SMVI with 100 and 500 components, respectively.}
	\label{fig:fwi_svgd_smvi_comparison_mean_std_error}
\end{figure}

Table \ref{table:fwi_cost} also indicates that the main factor controlling the difference in computational cost between SVGD and SMVI is the number of particles or components used in each algorithm. In addition, the above test shows that SMVI with 100 components provides results comparable to those obtained using SVGD with 500 particles. Below we investigate the influence of the number of particles or components on the performance of the 2 methods in two additional tests, one using SVGD with 100 particles and the other using SMVI with 500 components, so that for each method we can compare results with 100 and 500 particles (components), given the results above. The corresponding posterior mean, standard deviation, and relative error maps are displayed in Figure \ref{fig:fwi_svgd_smvi_comparison_mean_std_error}. Little obvious change is observed between Figures \ref{fig:fwi_svgd_smvi_comparison_mean_std_error}a and b, or between Figures \ref{fig:fwi_svgd_smvi_comparison_mean_std_error}c and d. This may indicate that 1D posterior statistics can already be represented reasonably well with only 100 model particles in SVGD or 100 components in SMVI. In this case, increasing the number of particles or components does not change these statistics significantly.

\begin{figure}
	\centering\includegraphics[width=\textwidth]{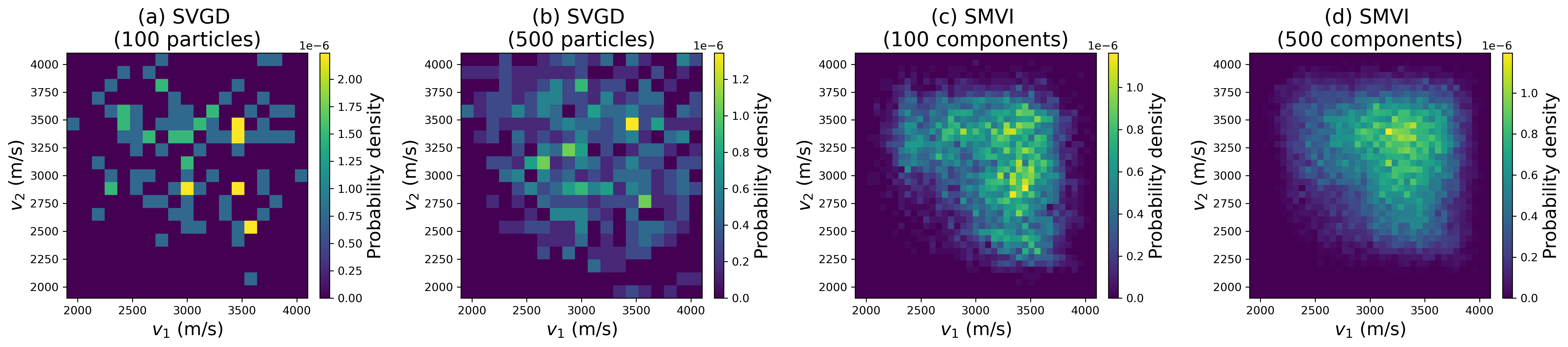}
	\caption{Posterior bivariate marginal distributions of velocity values at two vertically adjacent locations, (2.4 km, 0.84 km) and (2.4 km, 0.86 km). Their locations are marked by a red circle in Figure \ref{fig:fwi_vel_true_prior}a.}
	\label{fig:fwi_svgd_smvi_comparison_bi_2dmarginals}
\end{figure}

To support this interpretation and further examine the effect of the number of particles or components, posterior bivariate marginal distributions at (2.4 km, 0.84 km) and (2.4 km, 0.86 km) are shown in Figure \ref{fig:fwi_svgd_smvi_comparison_bi_2dmarginals}, corresponding to the same locations as in Figure \ref{fig:fwi_4methods_bi_2dmarginals}. In Figure \ref{fig:fwi_svgd_smvi_comparison_bi_2dmarginals}a, the 2D parameter space is sampled more sparsely than in panel b because of the smaller number of particles. In this case, 100 particles are used, corresponding to about 10 samples per dimension. With 500 particles, the SVGD approximation provides a denser representation of the 2D space, and the resulting distribution appears (slightly) better sampled. From this perspective, improved representation of posterior uncertainty, and thus improved inversion results, can be obtained in SVGD by increasing the number of particles. In contrast, each SMVI component has non-zero width and therefore contributes information over a neighbourhood of continuous parameter space rather than at a single point. In addition, as a parametric variational method, SMVI allows posterior samples to be generated easily from the mixture distribution after inversion \citep{zhao2026geopvi}, which can be used to describe posterior uncertainty more densely in parameter space. Consequently, in both Figures \ref{fig:fwi_svgd_smvi_comparison_bi_2dmarginals}c and \ref{fig:fwi_svgd_smvi_comparison_bi_2dmarginals}d, the 2D space is sampled densely and the difference between the two results is limited. 


\begin{figure}
	\centering\includegraphics[width=\textwidth]{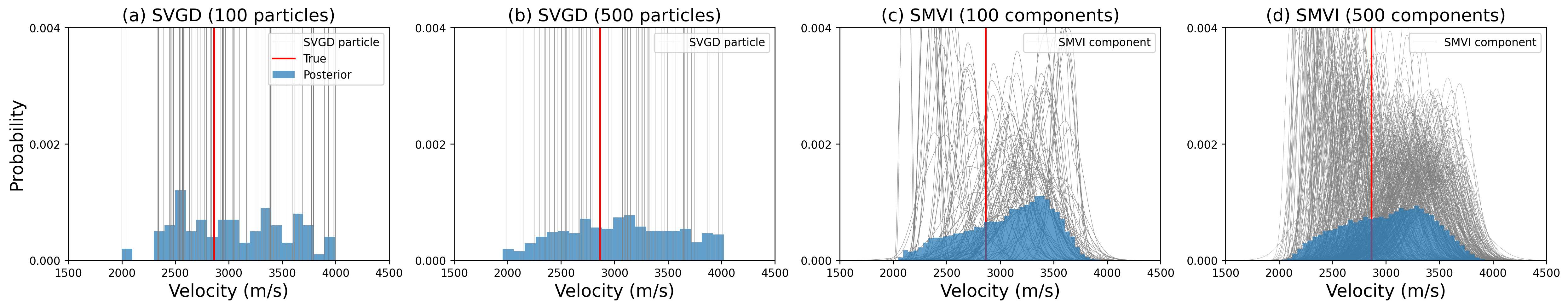}
	\caption{Posterior marginal distributions of velocity value at location (2.6 km, 0.9 km). In each panel, a vertical red line marks the true velocity value at this location, and a blue histogram displays the posterior pdf from each inversion test. Vertical grey lines in panels (a) and (b) denote velocity values from different SVGD particles, and grey histograms in panels (c) and (d) show component distributions from SMVI.}
	\label{fig:fwi_svgd_smvi_comparison_1Dmarginal}
\end{figure}

One possible reason why no obvious difference is observed between the SMVI results obtained using 100 and 500 components is that multiple components may overlap, making additional components redundant in the representation of the posterior pdf. To demonstrate this, posterior marginal pdf's at location (2.6 km, 0.9 km) are shown in Figure \ref{fig:fwi_svgd_smvi_comparison_1Dmarginal}. In each panel, the true velocity value is marked by a vertical red line, and the posterior pdf from each inversion test is displayed by a blue histogram. Similarly to Figure \ref{fig:1D_mixture_Gaussian}, grey vertical lines in Figures \ref{fig:fwi_svgd_smvi_comparison_1Dmarginal}a and b denote velocity values from different SVGD particles, whereas grey outlines in Figures \ref{fig:fwi_svgd_smvi_comparison_1Dmarginal}c and d represent component distributions from different SMVI components.

Since each SVGD particle represents only a point estimate and can be interpreted as a Delta function with zero width (equation \ref{eq:smvi_svgd}), different particles do not overlap. This again suggests that increasing the number of particles in SVGD improves the representation of the posterior uncertainty. In contrast, in panel d, most component distributions from SMVI overlap with one another. A similar overlapping behaviour can be observed in panel c, although less densely. 

\begin{figure}
	\centering\includegraphics[width=\textwidth]{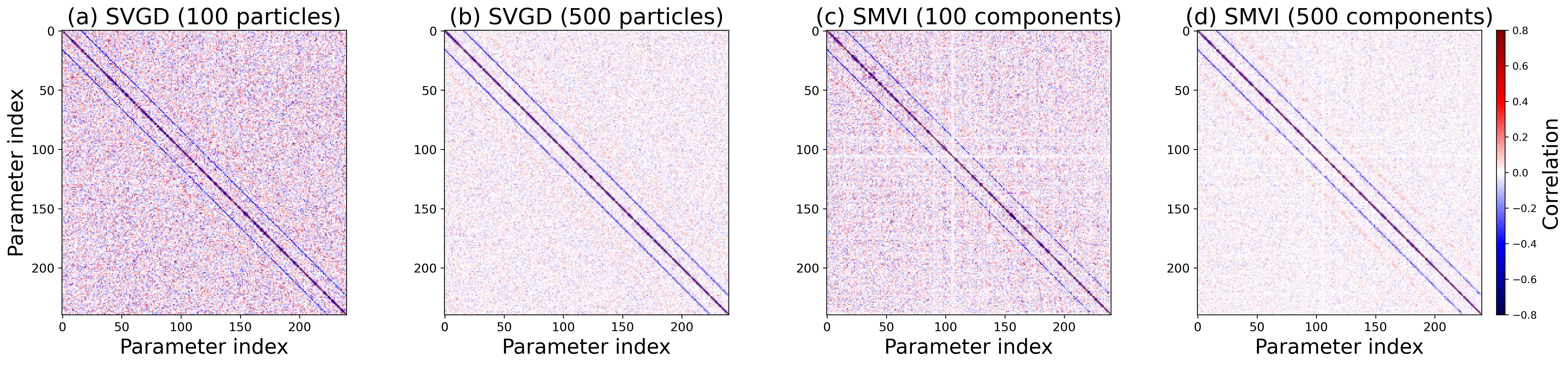}
	\caption{Posterior correlation matrices for velocity values inside a black box in Figure \ref{fig:fwi_vel_true_prior}a, calculated from different inversion tests indicated by the panel titles.}
	\label{fig:fwi_svgd_smvi_comparison_correlation}
\end{figure}

Finally, posterior correlation coefficients between model parameters estimated using different methods are compared. Figure \ref{fig:fwi_svgd_smvi_comparison_correlation} shows correlation matrices for velocity values within a black box in Figure \ref{fig:fwi_vel_true_prior}a, obtained from the 4 inversion tests. All 4 panels show off-diagonal inter-parameter correlation structures that are consistent with previous FWI studies \cite[e.g.,][]{zhao2024physically, zhao2024variational}. By comparing panels a, b with panels c, d, clearer off-diagonal correlation patterns with lower levels of background noise are observed when the number of particles in SVGD and the number of components in SMVI are increased, respectively.

Overall, this FWI example indicates that SMVI provides a more flexible, non-Gaussian posterior approximation than does ADVI or PSVI which both rely on Gaussian assumptions, particularly when the true posterior solution is multimodal. In addition, posterior uncertainty information is represented more densely and more accurately in the SMVI results than in the SVGD results, because each SMVI component represents a continuous distribution with non-zero width rather than a single point estimate. Given comparable inversion results, SMVI is therefore shown to be more computationally efficient than SVGD.

\subsection{Love wave tomography of the British Isles}
In a final example, SMVI is applied to real data by performing Love wave tomography of the British Isles. This problem has already been tackled using Monte Carlo methods \citep{galetti2017transdimensional}, providing a reference solution with which our results can be compared. 

\begin{figure}
	\centering\includegraphics[width=\textwidth]{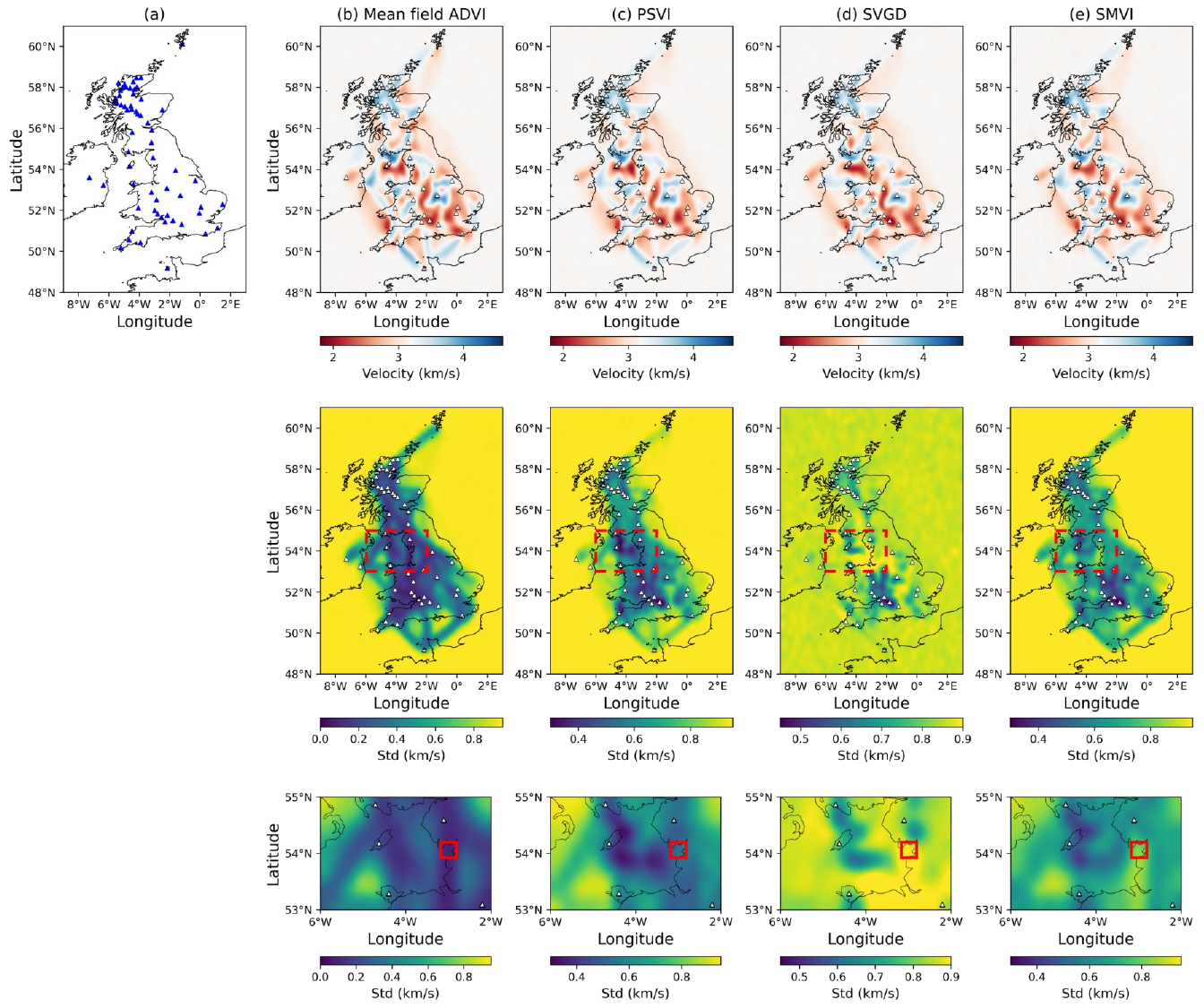}
	\caption{(a) Target area around the British Isles used in the Love wave tomography example. Locations of 61 seismometers are shown by blue triangles. (b)-(e) Posterior mean (top row) and standard deviation (middle row) maps of 10 s period Love wave tomography results for the British Isles obtained using different inversion methods indicated by the panel titles. White triangles show receiver locations. In the middle row, dashed red rectangles highlight regions around the East Irish Sea. A zoom of the corresponding standard deviation maps inside of the rectangles is shown in the bottom row in panels (b)-(e), in which loop-like high uncertainty structures surrounding a relatively lower uncertainty body are observed from the results using SVGD and SMVI. In the bottom row, red hollow squares mark a spatial location at which results from SVGD and SMVI show high uncertainty whereas results from ADVI and PSVI show low uncertainty. Posterior marginal distributions of Love wave velocity value at this location are compared in Figure \ref{fig:tomo_1dmarginals}.}
	\label{fig:tomo_mean_std}
\end{figure}

Figure \ref{fig:tomo_mean_std}a shows the locations of 61 seismometers (blue triangles) distributed around the British Isles. Ambient noise data from pairs of stations are cross-correlated to estimate Love wave propagation information at a period of 10 s. The resulting surface wave dispersion dataset is used to estimate Love wave velocity maps beneath the region. Further details of the data processing can be found in \cite{galetti2017transdimensional}. For inversion, the target region (Figure \ref{fig:tomo_mean_std}a) is discretised into a horizontal regular grid of 49 $\times$ 53 cells with a spacing of 0.25$^\circ$ in both directions, resulting in a dimensionality of 2597. The prior pdf for each velocity cell is defined as a uniform distribution bounded between 1.56 km/s and 4.81 km/s. Mean field ADVI, PSVI, SVGD and SMVI are applied to this tomography problem. In addition, a trans-dimensional Monte Carlo solution to this problem is available in \cite{galetti2015uncertainty} and \cite{galetti2017transdimensional}.

Figures \ref{fig:tomo_mean_std}b-\ref{fig:tomo_mean_std}e show the posterior mean (top row) and standard deviation (middle row) maps obtained using the 4 methods. As in the FWI example, the 4 posterior mean velocity maps are broadly similar and all delineate the main subsurface structures. These structures are consistent with seismic imaging results from previous studies \citep{hardwick2009new, nicolson2014rayleigh, zhao2022interrogating, bonadio2021optimal, zhao2025bayesian} and with the known geology of the region \cite[e.g.,][]{trewin2002geology, hardwick2009new, pharaoh2018anglo}.

In the middle row, the standard deviation maps differ between the 4 methods. Different colour scales are used in different panels to make the corresponding uncertainty structures visible. Nevertheless, the overall magnitude of the standard deviation values from mean field ADVI is much smaller than that from the other 3 methods, consistent with previous examples. The other three panels show similar spatial distributions of low and high uncertainty values in regions inside of the array where data sensitivity is high. As highlighted by red dashed rectangles in the middle row of Figures \ref{fig:tomo_mean_std}d and e, loop-like high uncertainty structures are found by SVGD and SMVI around the East Irish Sea, inside of which, relatively lower uncertainty values are observed. To further illustrate this feature, we display a zoom of this area in the bottom row in Figure \ref{fig:tomo_mean_std}. In the mean velocity maps, this region corresponds to consistently low velocity values associated with the East Irish Sea sedimentary basins \citep{galetti2017transdimensional}. Previously, \cite{galetti2015uncertainty} performed trans-dimensional reversible-jump Markov chain Monte Carlo (rj-McMC) inversion using this dataset, and showed that this high uncertainty loop arises because velocity models with different basin shapes and velocity values can fit the surface wave dataset within its uncertainty, leaving the boundaries of the corresponding region relatively under constrained. The similar uncertainty structures obtained using SVGD and SMVI therefore support the accuracy of the corresponding inversion results; the opposite (low uncertainty) results shown at the same location (red squares) in the lower panels of Figure \ref{fig:tomo_mean_std}a and b indicate that those results may be less accurate. 

\begin{figure}
	\centering\includegraphics[width=\textwidth]{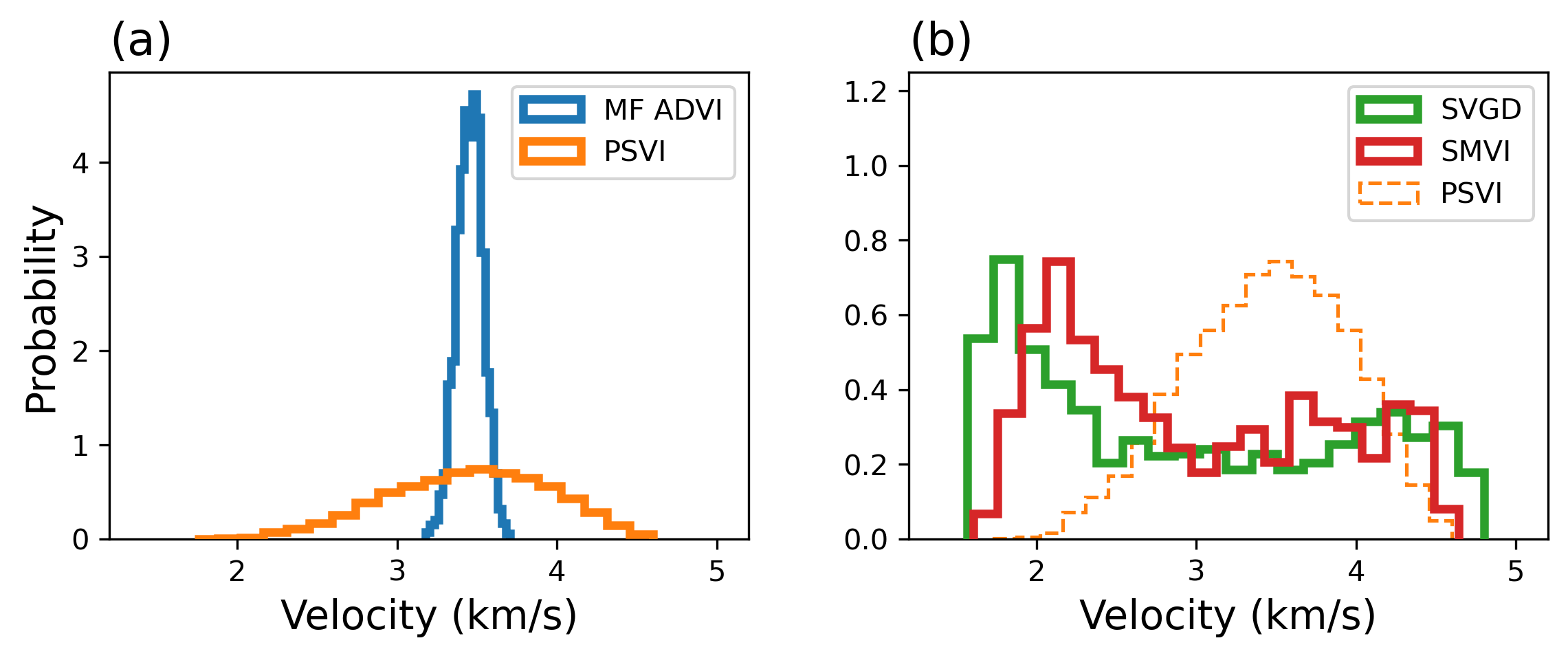}
	\caption{Posterior marginal distributions of Love wave velocity value at a spatial location marked by a red hollow square in the middle row of Figure \ref{fig:tomo_mean_std}e. Panel (a) shows histograms obtained using mean field ADVI (MF ADVI) and PSVI; panel (b) shows those obtained using SVGD, SMVI and PSVI. Note that different Y-axis scales are used in the two panels. The solid orange histogram in (a) and the dashed orange histogram in (b) represent the same marginal distribution.}
	\label{fig:tomo_1dmarginals}
\end{figure}

Posterior marginal distributions at a location marked by red hollow squares in the bottom row of Figure \ref{fig:tomo_mean_std} are shown in Figure \ref{fig:tomo_1dmarginals}. Figure \ref{fig:tomo_1dmarginals}a shows the results obtained using mean field ADVI and PSVI, and Figure \ref{fig:tomo_1dmarginals}b shows those obtained using SVGD and SMVI, with the PSVI result repeated for reference. Note that the Y-axis scales differ between the two panels. The posterior marginals obtained using mean field ADVI and PSVI are unimodal because both methods rely on transformed Gaussian posterior approximations, although the histogram from mean field ADVI is narrower than that from PSVI, since dominant parameter correlations are incorporated in PSVI. In contrast, posterior marginal distributions obtained using SVGD and SMVI are broader and bimodal. The distributions are peaked at both low and high velocity values, likely because velocity models with different boundaries of the East Irish Sea basins can explain the observed data almost equally well, so that this location may be classified as either inside or outside of the sedimentary basin. These bimodal distributions are consistent with the rj-McMC results presented in \cite{galetti2015uncertainty}, indicating that the posterior approximations obtained using SVGD and SMVI may be close to the reference Bayesian posterior solution.

\section{Discussion}
An important perspective is to regard the proposed SMVI framework as part of a progressive development of variational inference methods towards increasingly expressive posterior representations. At one extreme, \textit{maximum a posteriori} (MAP) estimation seeks only a single model that maximises the posterior probability density value, yielding a point estimate without quantifying its uncertainty. Particle-based methods such as SVGD replace the single MAP solution by a finite collection of particles whose empirical distribution approximates the posterior pdf, thereby allowing uncertainty quantification through a nonparametric particle representation. Parametric variational inference methods, such as ADVI and PSVI, instead replace a single point estimate from the MAP solution by a parametrised variational distribution that best approximates the posterior pdf within a chosen variational family, by minimising the KL-divergence. Compared with MAP estimation, these methods provide an explicit probabilistic description of posterior uncertainty while retaining a compact parametric representation.

From this perspective, the proposed SMVI framework can be viewed as a natural extension of parametric variational inference. Employing a mixture of variational components increases the flexibility of the variational family and allows more complex, multimodal, and non-Gaussian posterior distributions to be represented. As discussed in Section \ref{section:comparison}, this mixture representation also establishes a conceptual link between parametric and particle-based inference: a particle-based method can be interpreted as a limiting case in which each component collapses to a point mass, whereas conventional parametric variational inference corresponds to the single component case. To date, parametric and particle-based variational inference have been the two principal classes of variational methods applied successfully to geophysical inverse problems. In this sense, SMVI provides a unified framework that connects these two families while offering a compromise between posterior expressiveness and computational efficiency.

Another important observation from the numerical examples is that comparisons based only on posterior mean and standard deviation models may obscure meaningful differences between posterior solutions. In the FWI example, global posterior statistics obtained using PSVI, SVGD and SMVI are broadly similar, whereas local marginal and bivariate distributions reveal clearer differences in the obtained posterior pdf's. This suggests that assessment of Bayesian inversion methods should not rely solely on first- and second-order posterior summaries, particularly for nonlinear inverse problems in which multimodality and local non-Gaussian structure can be important \cite[see detailed comparison and discussion in][]{zhao2026linearised}.

In this study, each mixture component is represented by an uncorrelated transformed Gaussian distribution, corresponding to mean field ADVI. A natural extension would be to employ correlated Gaussian components, for example using the PSVI parametrisation, so that each component can represent local parameter correlations explicitly. A more expressive component distribution may reduce the number of mixture components required to approximate the target posterior pdf accurately, and can therefore reduce the number of forward simulations required by SMVI (Table \ref{table:fwi_cost}). The trade-off is that evaluation of the probability density of each component requires inversion of the corresponding covariance matrix, introducing additional computational overhead. Future work could therefore investigate the balance between these costs to determine whether the improved expressiveness of correlated components outweighs the increased cost of covariance matrix operations in large scale geophysical inverse problems. In addition, currently, different component distributions share the same weight coefficient. Introducing a scalar weight factor to each component in the mixture distribution can possibly also reduce the required number of components.

Despite the variety of open research areas that might be explored in future, the results herein already demonstrate that SMVI is capable of representing complex and multimodal solutions to probabilistic inverse problems. From an efficiency point of view, it offers tangible computational advantages over a close relative algorithm, BVI, which uses similar component mixtures to represent probability distributions, and which has been tested extensively in \cite{zhao2024bayesian,zhao2024physically}. And conceptually, SMVI bridges the gap between particle-based and parametric representations of posterior probability distributions, and between several associated algorithms that are used to estimate them.

\section{Conclusion}
In this paper, we introduced SMVI for Bayesian geophysical inversion with multimodal posterior solutions. The method combines parametric variational inference, such as in ADVI and PSVI, with optimisation methods similar to SVGD. In SMVI, the posterior distribution is approximated using a mixture variational model, in which each component distribution is parametrised by a set of parameters, called a particle in SMVI, and the component ensemble is optimised using the SVGD update rule. This provides a means to combine the computational efficiency of parametric variational inference with the flexibility of particle-based methods. The new method was tested using two seismic inverse problems. In a surface wave inversion example, posterior solutions obtained using SMVI were highly consistent with results from Markov chain Monte Carlo sampling and correctly represented a multimodal posterior solution. In a 2D acoustic full waveform inversion example, the SMVI approximation represented non-Gaussian posterior pdf more flexibly than ADVI and PSVI, while remaining more computationally efficient than SVGD and BVI. A real data Love wave tomography application around the British Isles further demonstrated the applicability of the method to practical seismic inversion. These results indicate that SMVI provides a practical and effective approach for Bayesian geophysical inversion when posterior distributions are complex and multimodal. More broadly, the method provides a flexible framework in which existing parametrised variational families can be incorporated into a mixture-based Stein variational formulation for efficient uncertainty quantification.

\begin{acknowledgments}
We thank the Edinburgh Imaging Project (EIP - \url{https://blogs.ed.ac.uk/imaging/}) sponsors (BP, TotalEnergies and Shell) for supporting this research. The authors thanks Sixtine Dromigny for contributory discussions about designing a multimodal Bayesian problem using an energy likelihood function. For the purpose of open access, we have applied a Creative Commons Attribution (CC BY) licence to any Author Accepted Manuscript version arising from this submission.
\end{acknowledgments}

\bibliographystyle{gji}
\bibliography{./reference}

\label{lastpage}
\end{document}